\documentclass[12pt]{article}

\usepackage{amsmath}
\usepackage{amsfonts}
\usepackage{amssymb}
\usepackage{cite}

\usepackage{bbold}

\usepackage{geometry}
\newcommand{\refs}{\Psi_0}
\newcommand\at[2]{\left.#1\right|_{#2}}

\begin{document}

\newpage
\setcounter{page}{0}

\begin{titlepage}

\begin{center}
{\Large \textbf{Algebraic Bethe ansatz for 19-vertex models
with upper triangular K-matrices}}\\
\vspace{1cm}
{\large
R.A. Pimenta\footnote{pimenta@df.ufscar.br} and
A. Lima-Santos\footnote{dals@df.ufscar.br}} \\
\vspace{1cm}
{\large
\textit{Universidade Federal de S\~ao Carlos, Departamento de F\'{\i}sica \\
Caixa Postal 676, CEP 13569-905, S\~ao Carlos, Brasil}
}\\
\end{center}
\vspace{1cm}

\begin{abstract}
By means of an algebraic Bethe ansatz approach we study the
Zamolodchikov-Fateev and Izergin-Korepin vertex models
with non-diagonal boundaries, characterized by reflection matrices with
an upper triangular form. Generalized Bethe vectors are used to diagonalize
the associated transfer matrix. The eigenvalues as well as the Bethe
equations are presented.
\end{abstract}

\vspace{1cm}
\begin{center}
Keywords: Algebraic Bethe Ansatz, Open boundary conditions
\end{center}
\vfill
\begin{center}
\small{\today}
\end{center}
\end{titlepage}

\newpage

\setcounter{footnote}{0}

\section{Introduction}\label{intro}

The study of integrable open vertex models in
the framework of the quantum inverse scattering
method was initiated in the seminal work of Sklyanin \cite{Skly}.
Using the so called reflection matrices \cite{Che}, Sklyanin
proposed a double-row transfer matrix from which integrable
spin Hamiltonians with boundary couplings can be obtained.
In addition to the Yang-Baxter equation, a set of compatibility
conditions on the reflection matrices has to be considered
in order to guarantee the commutativity of the double-row transfer matrices.

In principle, it should be possible 
to deal with the diagonalization problem of the double-row transfer matrix by means of
the algebraic Bethe ansatz technique (ABA) \cite{QISM}.
It turns out, however, that the applicability of the ABA
depends much on the structure of the reflection matrices. In fact,
whilst the case of diagonal $K-$matrices is well understood,
see for instance
\cite{Fan,Guan,LiYueHou,KurakLima,LiShi} and
references therein, the general situation is much more complicated and has been
a target of intense investigation over the last years. We can mention, for instance,
the progress achieved in the non-diagonal case \cite{cao,Doikou,CRM} by mapping the original
problem in an equivalent one with, at best, one diagonal and one triangular boundary matrix.

A generalization of the ABA for the case
where both reflection matrices have a triangular form
was recently proposed \cite{BCR}. We then extended this
work proposing a systematic method to deal
with transfer matrices possessing annihilation operators
in their expressions \cite{PL}. The key point in these works is
to consider a superposition of auxiliary Bethe states as
eigenstates of the double-row transfer matrix. The coefficients
of such linear combination can be fixed
by requiring the vanishing of extra unwanted terms in the ABA
analysis.

The purpose of this paper is to extend our recent work for 
$19-$vertex models with upper triangular
boundary matrices. We remember that the solution of $19-$vertex models
plays an essential role in the study of vertex models associated with higher rank symmetries, in both
periodic and open boundary conditions \cite{MM,MR,Lima,LiShi}. In this note, we
choose as representative of this class of vertex models
the Zamolodchikov-Fateev (ZF) \cite{ZF} and
Izergin-Korepin (IK) \cite{IK} models, for which non-diagonal reflection
matrices are know \cite{KIM,IOZ,FHLSY,LimaK19}.

We recall here
that the study of $19-$vertex models with diagonal
$K-$matrices in the ABA framework was started in \cite{Fan},
subsequently considered in \cite{LSY} and thereafter reviewed in \cite{KurakLima}.
These results complemented the previous one obtained by means of the analytical
Bethe ansatz \cite{MN1,YB}, providing the corresponding Bethe vectors of
the double-row transfer matrix. Very recently,
the spectrum of $19-$vertex models with non-diagonal $K-$matrices has
been reported in the literature \cite{NepoXXX,HCLYSW,CCSYW};
the construction of the respective eigenstates remains however
an open problem.

This work is organized as follows. In section \ref{models} we
remind some notation and basic ingredients necessary for the boundary
algebraic Bethe ansatz. We also give the $R-$matrix and
the upper tringular $K-$matrices expressions for the models we are
considering. In section \ref{BABA} we implement the
algebraic Bethe ansatz handling the first, second and third excited
states in detail. Our final remarks are given in section \ref{conclu} and
in the appendices necessary relations in the main text are given.

\section{The models and the triangular $K-$matrices}\label{models}

The fundamental object in a two-dimensional integrable theory is the
$R-$matrix satisfying the celebrated Yang-Baxter equation,
\begin{equation}\label{YB}
 R_{12}(u-v)R_{13}(u)R_{23}(v)=R_{23}(v)R_{13}(u)R_{12}(u-v),
\end{equation}
where the indices $ij$ in $R_{ij}$ indicate the spaces where
the $R-$matrix acts non-trivially and $u,v$ are called spectral
parameters. From the $R-$matrix, one can construct the
so called transfer matrix,
\begin{equation}\label{transfer}
 t_{\textrm{periodic}}(u)=\textrm{Tr}_a[R_{a1}(u)\ldots R_{aL}(u)],
\end{equation}
where $a$ denotes an auxiliary space and $n=1,2,\ldots,L$ refers
to a quantum vector space at the site $n$. In this context,
the Yang-Baxter equation is a sufficient
condition for the commutativity of transfer matrices
in periodic boundary conditions. The integrability follows
from the fact that the expansion of the transfer matrix in the
spectral parameter generates an infinite number of conserved
quantities.

The introduction of boundaries preserving integrability
can be performed through the introduction of
reflection matrices $K$ attached in each end of the chain \cite{Skly}.
Indeed, the $K-$matrices allow us to connect the single-row monodromy matrices
$T_{a}(u)=R_{a1}(u)\ldots R_{aL}(u)$ and $\hat{T}_a(u)\equiv T_{a}^{-1}(-u)$
in order define a double-row monodromy product,
\begin{equation}
U_{a}(u)=T_{a}(u)K^{-}(u)T_{a}^{-1}(-u)
\end{equation}
as well as the double-row transfer matrix,
\begin{equation}\label{t}
t(u)=\textrm{Tr}_{a}[K^{+}(u)T_{a}(u)K^{-}(u)\hat{T}_a(u)].
\end{equation}

The commutativity of the transfer matrix (\ref{t}) for arbitrary spectral
parameters is assured if the $K$-matrices
satisfy the reflection equations \cite{Skly,MN2},
\begin{equation}\label{REL}
 R_{12}(u-v)K_{1}^{-}(u)R_{21}(u+v)K_{2}^{-}(v)=K_{2}^{-}(v)R_{12}(u+v)K_{1}^{-}(u)R_{21}(u-v),
\end{equation}
\begin{eqnarray}\label{RER}
&&R_{12}(v-u)K_{1}^{+t_{1}}(u)M_1^{-1}R_{21}(-u-v-2\rho)M_1K_{2}^{+t_{2}}(v)\nonumber\\&=&
K_{2}^{+t_{2}}(v)M_1R_{12}(-u-v-2\rho)M_1^{-1}K_{1}^{+t_{1}}(u)R_{21}(v-u)
\end{eqnarray}
where $K^{+}$ represents the left boundary and $K^{-}$ the right one. The matrix $M$ and
the parameter $\rho$ are related to crossing-unitarity properties of the $R-$matrix
and will be given below. The symbol $t_i$ denotes
matrix transpostion in the space $i$. In addition, the equations (\ref{YB},\ref{REL},\ref{RER}) imply the global relations, 
\begin{equation}\label{global1}
\check{R}(u-v)T(u)\otimes T(v)=T(v)\otimes T(u)\check{R}(u-v)
\end{equation}
and 
\begin{equation}\label{global2}
R_{12}(u-v)U_{1}(u)R_{21}(u+v)U_{2}(v)=U_{2}(v)R_{12}(u+v)U_{1}(u)R_{21}(u-v),
\end{equation}
where $\check{R}(u)=PR(u)$.

The Zamolodchikov-Fateev \cite{ZF} and Izergin-Korepin \cite{IK} models are two of the
known integrable trigonometric $19-$vertex models (see \textit{e.g.} \cite{19PT} and references therein).
The former can be considered as a direct
generalization of the symmetric six-vertex model
while the latter is associated with the twisted affine algebra
$A_{2}^{(2)}$. For both models, the $R-$matrix has the following
common structure,
\begin{equation}
R(u)=\left(
\begin{array}{ccc|ccc|ccc}
a(u) & 0 & 0 & 0 & 0 & 0 & 0 & 0 & 0 \\
0 & b(u) & 0 & c(u) & 0 & 0 & 0 & 0 & 0 \\
0 & 0 & f(u) & 0 & d(u) & 0 & h(u) & 0 & 0 \\\hline
0 & c(u) & 0 & b(u) & 0 & 0 & 0 & 0 & 0 \\
0 & 0 & \tilde{d}(u) & 0 & e(u) & 0 & d(u) & 0 & 0 \\
0 & 0 & 0 & 0 & 0 & b(u) & 0 & c(u) & 0 \\\hline
0 & 0 & \tilde{h}(u) & 0 & \tilde{d}(u) & 0 & f(u) & 0 & 0 \\
0 & 0 & 0 & 0 & 0 & c(u) & 0 & b(u) & 0 \\
0 & 0 & 0 & 0 & 0 & 0 & 0 & 0 & a(u)
\end{array}
\right),
\end{equation}
where, for the ZF model we have,
\begin{eqnarray}\label{ZF}
a(u) &=& 1,~~~b(u) = \frac{\sinh(u)}{\sinh{(u+\eta)}},~~~
c(u) = \frac{\sinh(\eta)}{\sinh(u+\eta)},\nonumber\\
\tilde{d}(u) &=& d(u) = \frac{\sinh(\eta)\sinh(u)}{\sinh(u+\frac{\eta}{2})\sinh(u+\eta)},~~~
f(u) = \frac{\sinh(u-\frac{\eta}{2})\sinh(u)}{\sinh(u+\frac{\eta}{2})\sinh(u+\eta)},
\nonumber\\
e(u) &=& \frac{\cosh(\frac{\eta}{2}-u)\cosh(u+\eta)-\cosh(\frac{\eta}{2})}
{\sinh(u+\frac{\eta}{2})\sinh(u+\eta)},~~
\tilde{h}(u) = h(u) = \frac{\sinh(\eta)\sinh(\frac{\eta}{2})}
{\sinh(u+\frac{\eta}{2})\sinh(u+\eta)},\nonumber\\
\end{eqnarray}
and, for the IK solution,
\begin{eqnarray}\label{IK}
a(u) &=& 1,~~~b(u) = \frac{\sinh(u)}{\sinh{(u+\eta)}},~~~
c(u) = \frac{\sinh(\eta)}{\sinh(u+\eta)},\nonumber\\
d(u) &=& 
\frac{e^{\eta}\sinh(\eta)\sinh(u)}{\cosh(u+\frac{3\eta}{2})\sinh(u+\eta)},~~
\tilde{d}(u) = -e^{-2\eta}d(u),\nonumber\\
f(u) &=& \frac{\cosh(u+\frac{\eta}{2})\sinh(u)}{\cosh(u+\frac{3\eta}{2})\sinh(u+\eta)},~~
e(u) = \frac{\cosh(u-\frac{\eta}{2})\sinh(u+2\eta)-\cosh(\frac{\eta}{2})\sinh(\eta)}
{\cosh(u+\frac{3\eta}{2})\sinh(u+\eta)},\nonumber\\
h(u) &=& \frac{\cosh(u+\frac{3\eta}{2})\sinh(u+\eta)-
e^{2\eta}\cosh(u+\frac{\eta}{2})\sinh(u)}
{\cosh(u+\frac{3\eta}{2})\sinh(u+\eta)},\nonumber\\
\tilde{h}(u) &=& \frac{\cosh(u+\frac{3\eta}{2})\sinh(u+\eta)-
e^{-2\eta}\cosh(u+\frac{\eta}{2})\sinh(u)}
{\cosh(u+\frac{3\eta}{2})\sinh(u+\eta)}.
\end{eqnarray}
Some important properties of these $R-$matrices are,
\begin{enumerate}
 \item $PT$ symmetry: $R_{21}(u)=R_{12}^{t_1t_2}(u)$;
 \item unitarity: $R_{12}(u)R_{21}(-u)=\mathbb{1}\otimes \mathbb{1}$; and
 \item crossing-unitarity: $R_{12}^{t_1}(u)M_1R_{12}^{t_2}(-u-2\rho)M_1^{-1}=
 \zeta(u)\mathbb{1}\otimes \mathbb{1}$,
\end{enumerate}
where $\mathbb{1}$ is the $3\times 3$ identity matrix and $\zeta(u)$ is
a scalar function. Considering that the matrix $M$ satisfies $\left[R_{12}(u),M\otimes M\right]=0$,
the crossing-unitarity fix the parameter $\rho$ for each model. We can have
\begin{equation}\label{crossZF}
 M=\mathbb{1}~~\textrm{and}~~\rho=\eta,
\end{equation}
for the ZF model and
\begin{equation}\label{crossIK}
 M=\left(
\begin{array}{ccc}
e^{-2\eta} & 0 & 0\\
0 & 1 & 0\\
0 & 0 & e^{2\eta}
\end{array}
\right)~~\textrm{and}~~\rho=3\eta
\end{equation}
for the IK model.

The reflection equations (\ref{REL},\ref{RER}) have been
analyzed in \cite{KIM,IOZ,FHLSY,LimaK19}. Here we are interested
in the upper triangular solutions, which can be read from the type-I solutions
classified in \cite{LimaK19} and written as,
\begin{equation}\label{KM}
 K^{-}(u)=\left(
\begin{array}{ccc}
k_{11}^{-}(u) & k_{12}^{-}(u) & k_{13}^{-}(u)\\
0 & k_{22}^{-}(u) & k_{23}^{-}(u)\\
0 & 0 & k_{33}^{-}(u)\\
\end{array}
\right),
\end{equation}
with
\begin{eqnarray}
 k_{11}^{-}(u)&=&\sinh(u+\xi_{-})\sinh\left(u+\xi_{-}-\frac{\eta}{2}\right),~~~
 k_{22}^{-}(u)=\sinh(\xi_{-}-u)\sinh\left(u+\xi_{-}-\frac{\eta}{2}\right),
 \nonumber\\
 k_{33}^{-}(u)&=&\sinh(u-\xi_{-})\sinh\left(u-\xi_{-}+\frac{\eta}{2}\right),~~~
 k_{12}^{-}(u)=\beta_{-}\sinh (2 u)\sinh\left(u+\xi_{-}-\frac{\eta}{2}\right),
 \nonumber\\
 k_{23}^{-}(u)&=&\beta_{-}\sinh (2 u)\sinh\left(\xi_{-}-u\right),~~~
 k_{13}^{-}(u)=\left[\frac{\beta_{-}^2\sinh\left(\frac{\eta }{2}\right)}{\sinh (\eta )}\right]
 \sinh (2 u) \sinh \left(2u-\frac{\eta }{2}\right),\nonumber\\
 \end{eqnarray}
for the ZF model and
\begin{eqnarray}
 k_{11}^{-}(u)&=&k_{33}^{-}(u)=\sinh\left(u+\frac{3\eta}{4}+\epsilon\frac{i\pi}{4}\right)
 \cosh\left(u+\frac{3\eta}{4}-\epsilon\frac{i\pi}{4}\right),\nonumber\\
 k_{22}^{-}(u)&=&\sinh\left(-u+\frac{3\eta}{4}+\epsilon\frac{i\pi}{4}\right)
 \cosh\left(u+\frac{3\eta}{4}-\epsilon\frac{i\pi}{4}\right),\nonumber\\
 k_{12}^{-}(u)&=&\beta_{-}\sinh(2u)\cosh\left(u+\frac{3\eta}{4}-\epsilon\frac{i\pi}{4}\right),
 \nonumber\\
 k_{23}^{-}(u)&=&\beta_{-}\sinh(2u)e^{-\eta}\sinh\left(u+\frac{3\eta}{4}+\epsilon\frac{i\pi}{4}\right),
 \nonumber\\
 k_{13}^{-}(u)&=&
 \left[-\frac{\beta_{-}^2e^{-\eta}}{\cosh(\frac{\eta}{2})}\right]
 \sinh(2u)
 \cosh\left(u+\frac{\eta}{4}+\epsilon\frac{i\pi}{4}\right)
 \sinh\left(u+\frac{3\eta}{4}+\epsilon\frac{i\pi}{4}\right),
\end{eqnarray}
for the IK model. The respective $K^{+}-$matrices are obtained by,
\begin{equation}\label{KP}
 K^{+}(u)=\at{K^{-}(-u-\rho)M}{\left\{\xi_{-}\rightarrow\xi_{+},\beta_{-}\rightarrow\beta_{+}\right\}}
\end{equation}
where $M$ and $\rho$ are given by (\ref{crossZF},\ref{crossIK}).

In the above expressions, $\xi_{\pm}$ and $\beta_{\pm}$ are free constants
while $\epsilon=\pm 1$. We note that the upper triangular matrices of the ZF
model have two additional parameters $\xi_{\pm}$ compared
with the IK model.

\section{Algebraic Bethe ansatz}\label{BABA}

In this section we apply the algebraic Bethe ansatz
in order to handle the spectral problem for the transfer
matrix (\ref{t}). The first step to be carried out is the choice
of a representation for the single-row monodromy matrices $T_a(u)$, $\hat{T}_a(u)$
and, consequently, for $U_a(u)$.
In the case of $19-$vertex models,
a convenient one is given by the following $3\times 3$ matrices in the auxiliary space $a$
\cite{Fan,LSY,KurakLima},
\begin{equation}
 T_a(u)=\left(
\begin{array}{lll}
T_{11}(u) & T_{12}(u) & T_{13}(u) \\
T_{21}(u) & T_{22}(u) & T_{23}(u) \\
T_{31}(u) & T_{31}(u) & T_{33}(u)
\end{array}
\right),~~~\hat{T}_a(u)=\left(
\begin{array}{lll}
\hat{T}_{11}(u) & \hat{T}_{12}(u) & \hat{T}_{13}(u) \\
\hat{T}_{21}(u) & \hat{T}_{22}(u) & \hat{T}_{23}(u) \\
\hat{T}_{31}(u) & \hat{T}_{31}(u) & \hat{T}_{33}(u)
\end{array}
\right),
\end{equation}
with operator entries defined on the Hilbert space $\otimes_{i=1}^L\mathbb{C}^3$.
The double-row monodromy matrix can be thus written as,
\begin{equation}\label{mono}
 U_a(u)=\left(
\begin{array}{lll}
\mathcal{A}_1(u) & \mathcal{B}_1(u) & \mathcal{B}_2(u) \\
\mathcal{C}_1(u) & \mathcal{A}_2(u) & \mathcal{B}_3(u) \\
\mathcal{C}_2(u) & \mathcal{C}_3(u) & \mathcal{A}_3(u)
\end{array}
\right),
\end{equation}
where the operators $\mathcal{A}_j$, $\mathcal{B}_j$ and $\mathcal{C}_j$
are given in terms of $T_{ij}$ and $\hat{T}_{ij}$. This step is important since
it provides commutation relations
between the entries of (\ref{mono}) thanks to the relation (\ref{global2}).

We next have to consider a reference state as well as the corresponding
action of the $U-$operators on it. Due to the structure of the right
$K-$matrices (\ref{KM}), the pseudovacuum
defined by
\begin{equation}\label{refstate}
\Psi _{0}=\left( 
\begin{array}{c}
1 \\ 
0 \\
0
\end{array}
\right) _{(1)}\otimes \left( 
\begin{array}{c}
1 \\ 
0 \\
0
\end{array}
\right) _{(2)}\otimes \cdots \otimes \left( 
\begin{array}{c}
1 \\ 
0 \\
0
\end{array}
\right) _{(L)},
\end{equation}
turns out to be a good choice of reference state \cite{CRM}.
In fact, taking into account the exchange
relations between the matrix elements of $T(u)$ and $\hat{T}(u)$, provided
by equation (\ref{global1}), as well as the expressions
for $T_{ij}(u)\refs$ and $\hat{T}_{ij}(u)\refs$, we are able to calculate\footnote{For more
details on this calculation see for instance \cite{KurakLima}.},
\begin{equation}\label{diagonalaction}
 \mathcal{D}_{j}(u)\Psi_0=\Delta_j(u)\Psi_0~~\textrm{for}~~j=1,2,3
\end{equation}
and
\begin{equation}\label{destructionaction}
\mathcal{C}_{j}(u)\refs=0~~\textrm{for}~~j=1,2,3,
\end{equation}
where
\begin{eqnarray}
 \Delta_1(u)&=&k_{11}^{-}(u),~~~
 \Delta_2(u)=\left[k_{22}^{-}(u)-f_1(u)k_{11}^{-}(u)\right]b(u)^{2L},\nonumber\\
 \Delta_3(u)&=&\left[k_{33}^{-}(u)-f_4(u)k_{11}^{-}(u)-f_3(u)k_{22}^{-}(u)\right]f(u)^{2L}.
\end{eqnarray}
The shifted operators $\mathcal{D}_{j}(u)$ are conveniently defined by,
\begin{equation}\label{defD}
\mathcal{D}_1(u)=\mathcal{A}_1(u),~~\mathcal{D}_{2}(u)=\mathcal{A}_{2}(u)-f_1(u)\mathcal{A}_{1}(u),~~
 \mathcal{D}_{3}(u)=\mathcal{A}_{3}(u)-f_2(u)\mathcal{A}_{1}-f_3(u)\mathcal{D}_{2}(u),
\end{equation}
where the auxiliary functions $f$ are given by,
\begin{eqnarray}
f_1(u)&=&c(2u),~~
f_2(u)=\tilde{h}(2 u),~~
f_3(u)=\frac{c(2u)\left[\tilde{h}(2u)-1\right]}{c(2u)^2-e(2u)},
\nonumber\\
f_4(u)&=&\frac{c(2 u)^2-\tilde{h}(2 u)e(2u)}{c(2u)^2-e(2 u)}.
\end{eqnarray}

In terms of the operators (\ref{defD}), the transfer matrix
expression can be decomposed in two parts as follows,
\begin{equation}\label{transfer1}
 t(u)=t_d(u)+t_u(u)
\end{equation}
where
\begin{equation}\label{td}
 t_d(u)=\omega_1(u)\mathcal{D}_1(u)+\omega_2(u)\mathcal{D}_2(u)+\omega_3(u)\mathcal{D}_3(u)
\end{equation}
and,
\begin{equation}\label{tu}
 t_u(u)=k^{+}_{12}(u)\mathcal{C}_{1}(u)
   +k^{+}_{13}(u)\mathcal{C}_{2}(u)
   +k^{+}_{23}(u)\mathcal{C}_{3}(u).
\end{equation}
with
\begin{eqnarray}
\omega_1(u)&=&k_{11}^{+}(u)+f_1(u)k_{22}^{+}(u)+f_2(u)k_{33}^{+}(u),~~
\omega_2(u)=k_{22}^{+}(u)+f_3(u)k_{33}^{+}(u),\nonumber\\
\omega_3(u)&=&k_{33}^{+}(u).
\end{eqnarray}

Since the operators $\mathcal{C}_j(u)$ destroy the state (\ref{refstate}) we have
$t_u(u)\Psi_0=0$. Thus, the state (\ref{refstate}) is a common eigenstate
of the transfer matrices (\ref{transfer1}) and (\ref{td}) with eigenvalue,
\begin{equation}
 \Lambda_0(u)=\omega_1(u)\Delta_1(u)+\omega_2(u)\Delta_2(u)+\omega_3(u)\Delta_3(u).
\end{equation}

Within the ABA framework, the excited states of the term $t_d(u)$
can be generated by applying creation
operators on the reference state (\ref{refstate}) \cite{Fan,LSY,KurakLima}.
More precisely, the eigenstates of $t_d(u)$ are constructed by applying both $\mathcal{B}_1(u)$
and $\mathcal{B}_2(u)$ operators on $\Psi_0$. For instance,
the one-particle $\Psi_1(u_1)$ and the two-particle $\Psi_2(u_1,u_2)$ states are given by,
\begin{equation}
 \Psi_1(u_1)=\mathcal{B}_1(u_1)\Psi_0,
\end{equation}
\begin{equation}\label{PSI2}
 \Psi_2(u_1,u_2)=\mathcal{B}_1(u_1)\mathcal{B}_1(u_2)\Psi_0-\Gamma_2^{(2)}(u_1,u_2)\mathcal{B}_2(u_1)\Psi_0
\end{equation}
where the function $\Gamma_2^{(2)}(u_1,u_2)$
is fixed by requiring the symmetry
\begin{equation}\label{sympsi2}
\Psi_2(u_2,u_1)=\Omega(u_1,u_2)\Psi_2(u_1,u_2),~~~\Omega(u_1,u_2)=1/e_{01}(u_1,u_2)
\end{equation}
as well as the commutation relation (\ref{B1B1}).
Moreover, the multi-particle states satisfy a
Tarasov-like recurrence relation \cite{TARA}, namely,
\begin{eqnarray}\label{Psin}
 \Psi_n(u_1,\ldots,u_n)&=&\mathcal{B}_1(u_1)\Psi_{n-1}(u_2,\ldots,u_n)\nonumber\\&-&
 \mathcal{B}_2(u_1)\sum_{i=2}^n \Gamma_i^{(n)}(u_1,\ldots,u_n) \Psi_{n-2}(u_2,\ldots,\hat{u}_i,\ldots,u_n)
\end{eqnarray}
with the function $\Gamma_i$ given by,
\begin{eqnarray}
&&\Gamma_i^{(n)}(u_1,\ldots,u_n)= \prod_{j=2,j<i}^n \Omega(u_i,u_j)\nonumber\\&\times& 
 \left\{
 \Delta_1(u_i)e_{04}(u_1,u_i) \prod_{k=2,k\neq i}^n a_{11}(u_i,u_k)
 +
 \Delta_2(u_i)e_{05}(u_1,u_i) \prod_{k=2,k\neq i}^n a_{21}(u_i,u_k)
 \right\}.\nonumber\\
\end{eqnarray}
where, as usual, the notation $\hat{u}_i$ means that the rapidity
$u_i$ is absent in the function.

In our case,
however, the transfer matrix possesses the annihilation operators
$\mathcal{C}_{1,2,3}(u)$ in its expression and, as a result,
we have to seek for more intricate eigenstates. We propose that
a superposition of the states (\ref{Psin}) is needed to diagonalize
the full transfer matrix (\ref{transfer1}), namely,
\begin{equation}\label{phin}
 \Phi_n=\sum_{k=0}^{n} g^{(k)}\Psi_k.
\end{equation}
Therefore, the main task in this work is to fix the
formulas for the $g-$coefficients in (\ref{phin}). To this end,
we shall need to calculate the action of both $t_d(u)$ and
$t_u(u)$ on the vectors $\Psi_n$. 
Since the action of the shifted operators as well as
of the annihilation operators on the reference are known (\ref{diagonalaction},\ref{destructionaction}),
we will need to move the operators $\mathcal{D}_j$ and $\mathcal{C}_j$ over the
creation operators $\mathcal{B}_j$. This can be achieved by the repeated use of the
commutation relations given in appendix \ref{commutrela} which
are derived from equation (\ref{global2}).
After a very long though straightforward calculation, up to $n=3$, we are
able to propose that the action $t_d(u)\Psi_n$ can be written as,
\begin{eqnarray}\label{tdpsin}
 t_d(u)\Psi_n(u_1,\ldots,u_n) &=&
\Lambda_n(u,u_1,\ldots,u_n)\Psi_n(u_1,\ldots,u_n)\nonumber\\&+&
\mathcal{B}_1(u)\sum_{j=1}^n \mathcal{F}_j^{(n)}(u,u_1,\ldots,u_n)\Psi_{n-1}(u_1,\ldots,\hat{u}_j,\ldots,u_n)\nonumber\\&+&
\mathcal{B}_3(u)\sum_{j=1}^n \mathcal{G}_j^{(n)}(u,u_1,\ldots,u_n)\Psi_{n-1}(u_1,\ldots,\hat{u}_j,\ldots,u_n)\nonumber\\&+&
\mathcal{B}_2(u)\sum_{j<k}^n
\mathcal{H}_{jk}^{(n)}(u,u_1,\ldots,u_n)
\Psi_{n-2}(u_1,\ldots,\hat{u}_j,\ldots,\hat{u}_k,\ldots,u_n),\nonumber\\
\end{eqnarray}
while $t_u(u)\Psi_n$ has a more involved structure and it is
given by,
\begin{eqnarray}\label{tupsin}
 &&t_u(u)\Psi_n(u_1,\ldots,u_n) =
 \sum_{j=1}^n \mathcal{T}_j^{(n)}(u,u_1,\ldots,u_n) \Psi_{n-1}(u_1,\ldots,\hat{u}_j,\ldots,u_n)\nonumber\\&+&
 \sum_{j<k}^n \mathcal{U}_{jk}^{(n)}(u,u_1,\ldots,u_n)
 \Psi_{n-2}(u_1,\ldots,\hat{u}_j,\ldots,\hat{u}_k,\ldots,u_n)\nonumber\\&+&
 \mathcal{B}_1(u)
 \sum_{j<k}^n \mathcal{V}_{jk}^{(n)}(u,u_1,\ldots,u_n)\Psi_{n-2}(u_1,\ldots,\hat{u}_j,\ldots,\hat{u}_k,\ldots,u_n)
 \nonumber\\&+&
 \mathcal{B}_3(u)
 \sum_{j<k}^n \mathcal{W}_{jk}^{(n)}(u,u_1,\ldots,u_n)\Psi_{n-2}(u_1,\ldots,\hat{u}_j,\ldots,\hat{u}_k,\ldots,u_n)
 \nonumber\\&+&
  \mathcal{B}_1(u)
 \sum_{j<k<\ell}^n \mathcal{X}_{jk\ell}^{(n)}(u,u_1,\ldots,u_n)
 \Psi_{n-3}(u_1,\ldots,\hat{u}_j,\ldots,\hat{u}_k,\ldots,\hat{u}_{\ell},\ldots,u_n)
 \nonumber\\&+&
 \mathcal{B}_3(u)
 \sum_{j<k<\ell}^n \mathcal{Y}_{jk\ell}^{(n)}(u,u_1,\ldots,u_n)
 \Psi_{n-3}(u_1,\ldots,\hat{u}_j,\ldots,\hat{u}_k,\ldots,\hat{u}_{\ell},\ldots,u_n)
  \nonumber\\&+&
 \mathcal{B}_2(u)
 \sum_{j<k<\ell}^n \mathcal{Z}_{jk\ell}^{(n)}(u,u_1,\ldots,u_n)
 \Psi_{n-3}(u_1,\ldots,\hat{u}_j,\ldots,\hat{u}_k,\ldots,\hat{u}_{\ell},\ldots,u_n).
 \end{eqnarray}
where the functions $\mathcal{F},\mathcal{G},\ldots,\mathcal{Z}$ entering equations (\ref{tdpsin}) and (\ref{tupsin})
are given in appendix \ref{coetdtu}. We now use the above evaluations in order to consider
in detail the first, second and third excited states,
which allow us to propose a general formula for the 
$g-$coefficients.
 
\subsection{First excited state}

Considering our previous discussion we suppose as the first excited state of (\ref{transfer1}) the linear
combination,
\begin{equation}\label{ansatzPHI1}
 \Phi_1(u_1)=\Psi_1(u_1)+g(u_1)\Psi_0.
\end{equation}
with $g(u_1)$ to be fixed \textit{a posteriori}. Acting with $t(u)$ on it we obtain,
\begin{equation}
t(u)\Phi_1(u_1)=t_d(u)\Psi_1(u_1)+g(u_1)t_d(u)\Psi_0+t_u(u)\Psi_1(u_1),
\end{equation}
and, using $n=1$ in the formulas (\ref{tdpsin},\ref{tupsin}), we have,
\begin{eqnarray}\label{tPHI1}
t(u)\Phi_1(u_1)&=&\Lambda_1(u,u_1)\Phi_1(u_1)
+\mathcal{F}_1^{(1)}(u,u_1)\mathcal{B}_1(u)\Psi_0+\mathcal{G}_1^{(1)}(u,u_1)\mathcal{B}_3(u)\Psi_0\nonumber\\&+&
\left\{g(u_1)\left[\Lambda_0(u)-\Lambda_1(u,u_1)\right]+\mathcal{T}_1^{(1)}(u,u_1)\right\}\Psi_0
\end{eqnarray}
where
\begin{eqnarray}\label{lam1}
\Lambda_1(u,u_1)&=&\omega_1(u)\Delta _1(u)a_{11}(u,u_1)+
\omega_2(u)\Delta_2(u)(u)a_{21}(u,u_1)+\omega_3(u)\Delta_3(u)a_{31}(u,u_1).\nonumber\\
\end{eqnarray}

We see that $\Phi_1(u_1)$ will be an eigenstate of $t(u)$ with eigenvalue $\Lambda_1(u,u_1)$
provided that
the coefficients of the unwanted states $\mathcal{B}_1(u)\Psi_0$,
$\mathcal{B}_3(u)\Psi_0$ and $\refs$ vanish. The nullity requirement of
$\mathcal{F}_1^{(1)}(u,u_1)$
lead us to the constraint,
\begin{equation}\label{bethe1paux}
 \frac{\Delta_1(u_1)}{\Delta_2(u_1)}=-\frac{Q_2^{\mathcal{F}}(u,u_1)}{Q_1^{\mathcal{F}}(u,u_1)}
\end{equation}
and, due to the identity,
\begin{equation}
 \frac{Q_2^{\mathcal{F}}(u,u_1)}{Q_1^{\mathcal{F}}(u,u_1)}=
 \frac{Q_2^{\mathcal{G}}(u,u_1)}{Q_1^{\mathcal{G}}(u,u_1)}
\end{equation}
the function $\mathcal{G}_1^{(1)}(u,u_1)$ also vanishes if (\ref{bethe1paux})
is satisfied.
Considering the explicit expressions of the weights
we note that the dependence on the parameter $u$ disappears in the right-hand side
of (\ref{bethe1paux}) and,
as a result, we can write the Bethe equation constraining the rapidity $u_1$ as,
\begin{equation}\label{bethe1}
\frac{\Delta_1(u_1)}{\Delta_2(u_1)}=-\Theta(u_1)
\end{equation}
where, for the ZF model,
\begin{equation}
 \Theta(u_1)=\frac{\sinh(2u_1+\eta)\sinh\left(u_1+\xi_{+}+\frac{\eta}{2}\right)}
 {\sinh(2u_1)\sinh\left(u_1+\frac{\eta}{2}-\xi_{+}\right)}
\end{equation}
and, for the IK model,
\begin{equation}
 \Theta(u_1)=-\frac{\sinh(2u_1+\eta)\sinh\left(u_1+\frac{\eta}{4}-\epsilon\frac{3 i \pi}{4}\right)}
 {\sinh(2u_1)\sinh\left(u_1+\frac{3\eta}{4}-\epsilon\frac{i \pi}{4}\right)}
\end{equation}

We are left then with the coefficient of $\Psi_0$ in equation (\ref{tPHI1}). It can be used
to linearly extract an expression for $g(u_1)$,
\begin{equation}\label{expg}
 g(u_1)=\frac{\mathcal{T}_1^{(1)}(u,u_1)}{\Lambda_1(u,u_1)-\Lambda_0(u)}.
\end{equation}
To be an eigenstate of $t(u)$ the expression (\ref{ansatzPHI1}) cannot
depend on the spectral parameter $u$ and thus (\ref{expg}) seems
inconsistent. However, the dependence on $u$ disappears once we take into
account the Bethe equation (\ref{bethe1}). Indeed, the cumbersome expression (\ref{expg}),
if the rapidity $u_1$ is a solution of (\ref{bethe1}),
is simplified to
\begin{equation}\label{gfinalZF}
 g(u_1)=\beta_{+}\left[\frac{\sinh(2u_1+\eta)}
 {\sinh(u_1+\frac{\eta}{2}-\xi_{+})}\right]\Delta_2(u_1),
\end{equation}
in the case of the ZF model
and
\begin{equation}\label{gfinalIK}
 g(u_1)=\beta_{+}\left[\frac{\sinh(2u_1+\eta)}{\sinh(u_1+\frac{3\eta}{4}-\epsilon\frac{i\pi}{4})}\right]\Delta_2(u_1),
\end{equation}
for the IK model.

We have thus determined the first excited state. We observe that the transfer matrix $t(u)$
shares with $t_d(u)$ the eigenvalue expression (\ref{lam1})
as well as the Bethe equation (\ref{bethe1}).

\subsection{Second excited state}

For the second excited state the ansatz is,
\begin{eqnarray}\label{ansatzPHI2}
\Phi _{2}(u_{1},u_{2}) &=&
\Psi _{2}(u_{1},u_{2})\nonumber \\
&&
+g^{(1)}_2(u_{1},u_{2})\Psi _{1}(u_{1})+
g^{(1)}_{1}(u_{1},u_{2})\Psi_{1}(u_{2})  \nonumber \\
&&+
g^{(0)}_{12}(u_{1},u_{2})\Psi _{0}  
\end{eqnarray}
with the coefficients $g^{(1)}_{1,2}(u_{1},u_{2})$ and $
g^{(0)}_{12}(u_{1},u_{2})$ to be fixed in what follows.

As before,
we need to know the action of $t(u)$ on the state (\ref{ansatzPHI2}).
This can be done by setting $n=2$ in the expressions (\ref{tdpsin},\ref{tupsin}). As a result we
get the following off-shell expression,
\begin{eqnarray}\label{tPHI2}
 &&t(u)\Phi _{2}(u_{1},u_{2})=\Lambda_2(u,u_1,u_2)\Phi _{2}(u_{1},u_{2})\nonumber\\
 &+&
 \mathcal{F}_1^{(2)}(u,u_1,u_2)\mathcal{B}_1(u)\Psi_1(u_2)+
 \mathcal{F}_2^{(2)}(u,u_1,u_2)\mathcal{B}_1(u)\Psi_1(u_1)\nonumber\\
 &+&
 \mathcal{G}_1^{(2)}(u,u_1,u_2)\mathcal{B}_3(u)\Psi_1(u_2)+
 \mathcal{G}_2^{(2)}(u,u_1,u_2)\mathcal{B}_3(u)\Psi_1(u_1)\nonumber\\
 &+&
 \mathcal{H}_{12}^{(2)}(u,u_1,u_2)\mathcal{B}_2(u)\refs\nonumber\\
  &+&
  \left\{
  g_2^{(1)}(u_1,u_2)\mathcal{F}_1^{(1)}(u,u_1)+g_1^{(1)}(u_1,u_2)\mathcal{F}_1^{(1)}(u,u_2)+
  \mathcal{V}_{12}^{(2)}(u,u_1,u_2)
  \right\}\mathcal{B}_1(u)\refs
  \nonumber\\
  &+&
   \left\{
  g_2^{(1)}(u_1,u_2)\mathcal{G}_1^{(1)}(u,u_1)+g_1^{(1)}(u_1,u_2)\mathcal{G}_1^{(1)}(u,u_2)+
  \mathcal{W}_{12}^{(2)}(u,u_1,u_2)
  \right\}\mathcal{B}_3(u)\refs
  \nonumber\\
  &+&
\left\{g_1^{(1)}(u_1,u_2)\left[\Lambda_1(u,u_2)-\Lambda_2(u,u_1,u_2)\right]
+\mathcal{T}_1^{(2)}(u,u_1,u_2)
\right\}\Psi_1(u_2)
\nonumber\\
  &+&
\left\{g_2^{(1)}(u_1,u_2)\left[\Lambda_1(u,u_1)-\Lambda_2(u,u_1,u_2)\right]
+\mathcal{T}_2^{(2)}(u,u_1,u_2)
\right\}\Psi_1(u_1)
\nonumber\\
  &+&
\left\{
g_{12}^{(0)}(u_1,u_2)\right.\left[\Lambda_0(u)-\Lambda_2(u,u_1,u_2)\right]\nonumber\\&+&
g_1^{(1)}(u_1,u_2)\mathcal{T}_{1}^{(1)}(u,u_2)+
g_2^{(1)}(u_1,u_2)\mathcal{T}_{1}^{(1)}(u,u_1)+
\left.\mathcal{U}_{12}^{(2)}(u,u_1,u_2)
\right\}\refs
 \end{eqnarray}

We observe that the first five unwanted terms in (\ref{tPHI2})
coincide with the unwanted terms of
the action of the diagonal
transfer matrix $t_d(u)$ on the state $\Psi_2$ (\ref{PSI2}). Therefore, their vanishing lead to the
Bethe equations for the rapidities $u_1$ and $u_2$, namely
\begin{equation}\label{bethe2p}
\frac{\Delta_1(u_j)}{\Delta_2(u_j)}=
-\Theta(u_j)\prod_{k=1,k\neq j}^2 \frac{a_{21}(u_j,u_k)}{a_{11}(u_j,u_k)}~~\textrm{for}~~j=1,2.
\end{equation}
We remark that the constraints (\ref{bethe2p}) are obtained from the
equations $\mathcal{F}_j^{(2)}(u,u_1,u_2)=0$ taking into account
the explicit formulas for the Boltzmann weights. Due to
identities provenient from the Yang-Baxter and reflection algebras,
we note that the coefficients $\mathcal{G}_j^{(2)}(u,u_1,u_2)$
and $\mathcal{H}_{12}^{(2)}(u,u_1,u_2)$ also vanish
if (\ref{bethe2p}) are valid.

The other coefficients are used in order to extract the expressions
for the unknown functions in the ansatz state (\ref{ansatzPHI2}). The vanishing
requirement of the
coefficients of $\Psi_1(u_1)$, $\Psi_1(u_2)$ and $\refs$ allows us to write,
\begin{equation}\label{expg2paux1}
g_1^{(1)}(u_1,u_2)=\frac{\mathcal{T}_1^{(2)}(u,u_1,u_2)}{\Lambda_2(u,u_1,u_2)-\Lambda_1(u,u_2)},~~~
g_2^{(1)}(u_1,u_2)=\frac{\mathcal{T}_2^{(2)}(u,u_1,u_2)}{\Lambda_2(u,u_1,u_2)-\Lambda_1(u,u_1)}
\end{equation}
and
\begin{equation}\label{expg2paux2}
 g_{12}^{(0)}(u_1,u_2)=\frac{g_1^{(1)}(u_1,u_2)\mathcal{T}_{1}^{(1)}(u,u_2)+
g_2^{(1)}(u_1,u_2)\mathcal{T}_{1}^{(1)}(u,u_1)+
\mathcal{U}_{12}^{(2)}(u,u_1,u_2)
}{\Lambda_2(u,u_1,u_2)-\Lambda_0(u)}.
\end{equation}
Similarly to the first
excited state, the resulting expressions for coefficients (\ref{expg2paux1},\ref{expg2paux2})
contain the spectral parameter
$u$. Once again, this situation can be overcome if we consider the expression
for the ratio $\Delta_1/\Delta_2$ constrained by the Bethe equations (\ref{bethe2p}).
After some cumbersome manipulation we find that the expressions (\ref{expg2paux1},\ref{expg2paux2})
simplify to the following factorized
structure,
\begin{eqnarray}\label{expg1g2g3}
g_{1}^{(1)}(u_{1},u_{2}) &=&g(u_{1})a_{21}(u_1,u_2),~~  \nonumber \\
g_{2}^{(1)}(u_{1},u_{2}) &=&g(u_{2})a_{21}(u_{2},u_{1})\Omega(u_2,u_1),~~  \nonumber \\
g_{12}^{(0)}(u_{1},u_{2}) &=&g(u_{1})g(u_{2})s(u_1,u_2)
\end{eqnarray}
where $g(u_i)$ are the functions (\ref{gfinalZF}) or (\ref{gfinalIK}) provenient from
the first excited state analysis and,
\begin{equation}
 s(u_1,u_2)=\frac{\sinh(u_1+u_2)\sinh(u_1-u_2-\eta)\sinh\left(u_1+u_2+\frac{3\eta}{2}\right)}
 {\sinh\left(u_1-u_2-\frac{\eta}{2}\right)\sinh\left(u_1+u_2+\frac{\eta}{2}\right)^2}
\end{equation}
for the ZF model and
\begin{eqnarray}
 s(u_1,u_2)&=&\frac{\cosh\left(u_1+u_2+\frac{\eta}{2}\right)\sinh(u_1+u_2+2\eta)
 }
 {\sinh(u_1+u_2)\cosh\left(u_1-u_2+\frac{\eta}{2}\right)
 \cosh\left(u_1+u_2+\frac{3\eta}{2}\right)^2}\nonumber\\&\times&
 \left[
 \cosh\left(u_1+\frac{\eta}{4}-i\epsilon\frac{\pi}{4}\right)\right.
 \cosh\left(u_1+\frac{3\eta}{4}+i\epsilon\frac{\pi}{4}\right)\nonumber\\&&+
\left. \cosh\left(u_2-\frac{\eta}{4}+i\epsilon\frac{\pi}{4}\right)
 \cosh\left(u_2+\frac{5\eta}{4}-i\epsilon\frac{\pi}{4}\right)
 \right]
\end{eqnarray}
for the IK solution. We observe the appearing of the
factor $\Omega(u_1,u_2)$ in the coefficient $g_{2}^{(1)}(u_{1},u_{2})$.
As a consequence, we also have the symmetry property (\ref{sympsi2})
for the generalized state $\Phi_2(u_1,u_2)$.

At this point, we use the expressions (\ref{expg1g2g3}) in the
remaining unwanted terms of (\ref{tPHI2}) and, taking into account
the constrains on $u_1$ and $u_2$ dictated by the Bethe equations (\ref{bethe3p}),
we can check by direct computation that all of them are automatically cancelled.
In this way, we conclude that the state (\ref{ansatzPHI2}) with coefficients
(\ref{expg1g2g3}) is an eigenstate of (\ref{transfer1}) with eigenvalue $\Lambda_2(u,u_1,u_2)$ (\ref{lam}).

\subsection{Third excited state}

We proceed in a similar way for the third excited state. The ansatz now
is given by,
\begin{eqnarray}\label{phi3}
\Phi _{3}(u_{1},u_{2},u_{3}) &=&\Psi _{3}(u_{1},u_{2},u_{3})
+g_{3}^{(2)}(u_{1},u_{2},u_{3})\Psi _{2}(u_{1},u_{2})  \nonumber \\
&+& g_{2}^{(2)}(u_{1},u_{2},u_{3})\Psi_{2}(u_{1},u_{3})+
g_{1}^{(2)}(u_{1},u_{2},u_{3})\Psi _{2}(u_{2},u_{3}) \nonumber \\
&+&g_{23}^{(1)}(u_{1},u_{2},u_{3})\Psi_{1}(u_{1})+
g_{13}^{(1)}(u_{1},u_{2},u_{3})\Psi_{1}(u_{2})+
g_{12}^{(1)}(u_{1},u_{2},u_{3})\Psi _{1}(u_{3})  \nonumber \\
&+&g_{123}^{(0)}(u_{1},u_{2},u_{3})\Psi _{0}  
\end{eqnarray}
where the coefficients $g^{(k)}(u_{1},u_{2},u_{3})$ will be determined in
the forthcoming analysis.

Once again, we need to evaluate the action of $t(u)$ on the ansatz state (\ref{phi3}).
For this end we use $n=3$ in the expansions (\ref{tdpsin},\ref{tupsin}) and,
as a result, we note the appearing of many cumbersome unwanted terms. For this reason,
we omit their explicit formulas here. It turns out that the Bethe equations can be obtained
from the coefficients of the unwanted states $\mathcal{B}_1(u)\Psi_2(u_j,u_k)$, and it read
\begin{equation}\label{bethe3p}
\frac{\Delta_1(u_j)}{\Delta_2(u_j)}=
-\Theta(u_j)\prod_{k=1,k\neq j}^3 \frac{a_{21}(u_j,u_k)}{a_{11}(u_j,u_k)}~~\textrm{for}~~j=1,2,3.
\end{equation}
We then choose the simplest unwanted terms to obtain the $g-$functions.
For the amplitudes of order two, namely, $g^{(2)}(u_1,u_2,u_3)$,
we use the coefficients of $\Psi_2(u_j,u_k)$ and, after taking
into account the Bethe equations (\ref{bethe3p}), we find,
\begin{eqnarray}\label{g2}
 g_3^{(2)}(u_1,u_2,u_3)&=&
 g(u_3)a_{21}(u_3,u_1)a_{21}(u_3,u_2)\Omega(u_3,u_1)\Omega(u_3,u_2),\nonumber\\
 g_2^{(2)}(u_1,u_2,u_3)&=&
 g(u_2)a_{21}(u_2,u_1)a_{21}(u_2,u_3)\Omega(u_2,u_1),\nonumber\\
 g_1^{(2)}(u_1,u_2,u_3)&=&
 g(u_1)a_{21}(u_1,u_2)a_{21}(u_1,u_3).
\end{eqnarray}
The functions $g^{(1)}(u_1,u_2,u_3)$ are derived from the coefficients
of $\Psi_1(u_j)$ while $g^{(0)}(u_1,u_2,u_3)$ is obtained from the coefficient
of $\Psi_0$. Using the Bethe equations (\ref{bethe3p}), they acquire the following
structure,
\begin{eqnarray}\label{g1}
 g_{23}^{(1)}(u_1,u_2,u_3)&=&
 g(u_2)g(u_3)a_{21}(u_2,u_1)a_{21}(u_3,u_1)s(u_2,u_3)\Omega(u_2,u_1)\Omega(u_3,u_1),\nonumber\\
 g_{13}^{(1)}(u_1,u_2,u_3)&=&
 g(u_1)g(u_3)a_{21}(u_1,u_2)a_{21}(u_3,u_2)s(u_1,u_3)\Omega(u_3,u_2),\nonumber\\
 g_{12}^{(1)}(u_1,u_2,u_3)&=&
 g(u_1)g(u_2)a_{21}(u_2,u_3)a_{21}(u_1,u_3)s(u_1,u_2),\nonumber\\
 g_{123}^{(0)}(u_1,u_2,u_3)&=&
 g(u_1)g(u_2)g(u_3)s(u_1,u_2)s(u_1,u_3)s(u_2,u_3).
 \end{eqnarray}
Finally, we can see by direct computation that all the other unwanted terms
are cancelled provided that we take into account the expressions (\ref{g2},\ref{g1}) in
addition to the Bethe equations (\ref{bethe3p}). Therefore, the vector (\ref{phi3})
is an eigenstate of the transfer matrix (\ref{transfer1}) with energy $\Lambda_3(u,u_1,u_2,u_3)$
(\ref{lam}).

\subsection{General excited state}

Considering the results of the previous subsections we
are able to propose a solution of the spectral
problem associated with the transfer matrix (\ref{transfer1}):
the nth excited
eigenstate is given by 
\begin{eqnarray}\label{Phin}
&&\Phi _{n}(u_{1},\dots ,u_{n})=
\Psi_n(u_{1},\dots ,u_{n})  \nonumber \\
&+& \sum_{k=0}^{n-1} \sum_{\ell _{1}<\dots <\ell_{n-k}=1}^{n}
g^{(k)}_{\ell_{1},\dots ,\ell _{n-k}}(u_{1},\dots ,u_{n})
\Psi _{k}(u_{1},\ldots ,\hat{u}_{\ell _{1}},\ldots,\hat{u}_{\ell _{n-k}},\ldots ,u_{n}),
\end{eqnarray}
where the $\Psi_n$ vectors are obtained by the recurrence relation (\ref{Psin}) and
the $g-$functions have the following expression, 
\begin{equation}
g^{(k)}_{\ell _{1},\dots ,\ell _{n-k}}(u_{1},\dots ,u_{n})
= \prod_{m\in\bar{\ell}} g(u_{m})
\prod_{m^{\prime }\in\bar{\ell},m^{\prime }<m}s(u_{m^{\prime }},u_{m})
\prod_{m^{\prime \prime }=1,m^{\prime \prime}\notin \bar{\ell}}^{n} a_{21}(u_{m},u_{m^{\prime\prime}})
\tilde{\Omega}_{m,m^{\prime\prime}}
\end{equation}
with $\bar{\ell}=\{\ell _{1},\dots ,\ell _{n-k}\}$ and
\begin{equation}
\tilde{\Omega}_{m,m^{\prime\prime}}=
\left\{
\begin{array}{ll}
\Omega(u_m,u_{m^{\prime\prime}}), & ~~ \textrm{if $m>m^{\prime\prime}$}\\
1, & ~~ \textrm{otherwise.}
\end{array} \right.
\end{equation}
The
associated eigenvalue is given by,
\begin{equation}\label{lamfinal}
 \Lambda_n(u,u_1,\ldots,u_n)=\sum_{\alpha=1}^3\omega_\alpha(u)\Delta_{\alpha}(u)
 \prod_{j=1}^{n}a_{\alpha 1}(u,u_{j}),
\end{equation}
while the Bethe rapidities have to satisfy,
\begin{equation}\label{bethefinal}
\frac{\Delta_1(u_j)}{\Delta_2(u_j)}=
-\Theta(u_j)\prod_{k=1,k\neq j}^n \frac{a_{21}(u_j,u_k)}{a_{11}(u_j,u_k)},
\end{equation}
for $j=1,\ldots,n$.

\section{Conclusion} \label{conclu}

By means of a generalized Bethe
ansatz approach, we have presented a solution of the open Zamolodchikov-Fateev and
Izergin-Korepin vertex models with triangular boundaries. Remarkably, the eigenvalues of
the corresponding transfer matrix (\ref{lamfinal}) as well as the Bethe equations (\ref{bethefinal})
are exactly the same of the diagonal boundary case. On the other hand,
the eigenstates have to be constructed from a linear superposition (\ref{Phin})
of auxiliary Bethe states (\ref{Psin}). These results show that such structure is
not restricted to the six-vertex model \cite{BCR,PL} and could
be extended for open vertex models associated with higher rank algebras.

Our results may also be useful in the study of $19-$vertex models with
complete boundary matrices in the framework of the ABA.
More specifically, it would be interesting to generalize the
approach of the recent work \cite{BC} at least for rational versions of the models
considered here.

Another interesting further direction of investigation is a thorough analysis 
of the off-shell structure dictated by equations (\ref{tdpsin},\ref{tupsin}).
In fact, the quasi-classical limit of the ABA solution,
along the lines of \cite{ALS1}, can possibly lead to generalized
three-state Gaudin magnets and Knizhnik-Zamolodchikov equations.

\section{Acknowledgments}

We would like to thank Samuel Belliard for reading the manuscript
and for his comments. The work of RAP has been supported by S\~{a}o Paulo Research
Foundation (FAPESP), grant \#2012/13126-0. ALS thanks Brazilian
Research Council (CNPq), grant \#304054/ 2009-7, and FAPESP, grant
\#2011/18729-1, for financial support.

\appendix

\section{Commutation relations}
\label{commutrela}
\setcounter{equation}{0}
\renewcommand{\theequation}{A.\arabic{equation}}

In this section we present exchange relations coming
from equation (\ref{global2}) which are necessary to
evaluate (\ref{tdpsin}) and (\ref{tupsin}). We have,
\begin{eqnarray}\label{D1B1}
\mathcal{D}_{1}(u)\mathcal{B}_{1}(v) &=&  
  a_{11}(u,v)\mathcal{B}_{1}(v)\mathcal{D}_{1}(u) + a_{12}(u,v)\mathcal{B}_{1}(u)\mathcal{D}_{1}(v) + 
   a_{13}(u,v)\mathcal{B}_{1}(u)\mathcal{D}_{2}(v) \nonumber\\&+& a_{14}(u,v)\mathcal{B}_{2}(u)\mathcal{C}_{1}(v) + 
   a_{15}(u,v)\mathcal{B}_{2}(u)\mathcal{C}_{3}(v) + a_{16}(u,v)\mathcal{B}_{2}(v)\mathcal{C}_{1}(u),\nonumber\\
\end{eqnarray}
\begin{eqnarray}\label{D2B1}
\mathcal{D}_{2}(u)\mathcal{B}_{1}(v) &=&  
  a_{21}(u,v)\mathcal{B}_{1}(v)\mathcal{D}_{2}(u) + a_{22}(u,v)\mathcal{B}_{1}(u)\mathcal{D}_{1}(v) + 
   a_{23}(u,v)\mathcal{B}_{1}(u)\mathcal{D}_{2}(v) \nonumber\\&+& a_{24}(u,v)\mathcal{B}_{3}(u)\mathcal{D}_{1}(v) + 
   a_{25}(u,v)\mathcal{B}_{3}(u)\mathcal{D}_{2}(v) + a_{26}(u,v)\mathcal{B}_{2}(u)\mathcal{C}_{1}(v) \nonumber\\&+& 
   a_{27}(u,v)\mathcal{B}_{2}(u)\mathcal{C}_{3}(v) + a_{28}(u,v)\mathcal{B}_{2}(v)\mathcal{C}_{1}(u) + 
   a_{29}(u,v)\mathcal{B}_{2}(v)\mathcal{C}_{3}(u) \nonumber\\&+& a_{210}(u,v)\mathcal{B}_{1}(v)\mathcal{D}_{1}(u),
\end{eqnarray}
\begin{eqnarray}\label{D3B1}
\mathcal{D}_{3}(u)\mathcal{B}_{1}(v) &=&  
  a_{31}(u,v)\mathcal{B}_{1}(v)\mathcal{D}_{3}(u) + a_{32}(u,v)\mathcal{B}_{1}(u)\mathcal{D}_{1}(v) + 
   a_{33}(u,v)\mathcal{B}_{1}(u)\mathcal{D}_{2}(v) \nonumber\\&+& a_{34}(u,v)\mathcal{B}_{3}(u)\mathcal{D}_{1}(v) + 
   a_{35}(u,v)\mathcal{B}_{3}(u)\mathcal{D}_{2}(v) + a_{36}(u,v)\mathcal{B}_{2}(u)\mathcal{C}_{1}(v) \nonumber\\&+& 
   a_{37}(u,v)\mathcal{B}_{2}(u)\mathcal{C}_{3}(v) + a_{38}(u,v)\mathcal{B}_{2}(v)\mathcal{C}_{1}(u) + 
   a_{39}(u,v)\mathcal{B}_{2}(v)\mathcal{C}_{3}(u) \nonumber\\&+& a_{310}(u,v)\mathcal{B}_{1}(v)\mathcal{D}_{1}(u) + 
   a_{311}(u,v)\mathcal{B}_{1}(v)\mathcal{D}_{2}(u),
\end{eqnarray}
\begin{eqnarray}\label{C1B1}
\mathcal{C}_{1}(u)\mathcal{B}_{1}(v) &=&  
  c_{11}(u,v)\mathcal{B}_{1}(v)\mathcal{C}_{1}(u) + c_{12}(u,v)\mathcal{B}_{1}(v)\mathcal{C}_{3}(u) + 
   c_{13}(u,v)\mathcal{B}_{1}(u)\mathcal{C}_{3}(v) \nonumber\\&+& c_{14}(u,v)\mathcal{B}_{3}(u)\mathcal{C}_{3}(v) + 
   c_{15}(u,v)\mathcal{B}_{2}(v)\mathcal{C}_{2}(u) + c_{16}(u,v)\mathcal{D}_{1}(v)\mathcal{D}_{1}(u) \nonumber\\&+& 
   c_{17}(u,v)\mathcal{D}_{1}(v)\mathcal{D}_{2}(u) + c_{18}(u,v)\mathcal{D}_{1}(u)\mathcal{D}_{1}(v) + 
   c_{19}(u,v)\mathcal{D}_{1}(u)\mathcal{D}_{2}(v) \nonumber\\&+& c_{110}(u,v)\mathcal{D}_{2}(u)\mathcal{D}_{1}(v) + 
   c_{111}(u,v)\mathcal{D}_{2}(u)\mathcal{D}_{2}(v),
\end{eqnarray}
\begin{eqnarray}\label{C2B1}
\mathcal{C}_{2}(u)\mathcal{B}_{1}(v) &=&  
  y_{11}(u,v)\mathcal{B}_{1}(v)\mathcal{C}_{2}(u) + y_{12}(u,v)\mathcal{B}_{3}(v)\mathcal{C}_{2}(u) + 
   y_{13}(u,v)\mathcal{B}_{1}(u)\mathcal{C}_{2}(v) \nonumber\\&+& y_{14}(u,v)\mathcal{B}_{3}(u)\mathcal{C}_{2}(v) + 
   y_{15}(u,v)\mathcal{D}_{1}(v)\mathcal{C}_{1}(u) + y_{16}(u,v)\mathcal{D}_{2}(v)\mathcal{C}_{1}(u) \nonumber\\&+& 
   y_{17}(u,v)\mathcal{D}_{1}(v)\mathcal{C}_{3}(u) + y_{18}(u,v)\mathcal{D}_{2}(v)\mathcal{C}_{3}(u) + 
   y_{19}(u,v)\mathcal{D}_{1}(u)\mathcal{C}_{1}(v) \nonumber\\&+& y_{110}(u,v)\mathcal{D}_{2}(u)\mathcal{C}_{1}(v) + 
   y_{111}(u,v)\mathcal{D}_{3}(u)\mathcal{C}_{1}(v) + y_{112}(u,v)\mathcal{D}_{1}(u)\mathcal{C}_{3}(v) \nonumber\\&+& 
   y_{113}(u,v)\mathcal{D}_{2}(u)\mathcal{C}_{3}(v) + y_{114}(u,v)\mathcal{D}_{3}(u)\mathcal{C}_{3}(v),
\end{eqnarray}
\begin{eqnarray}\label{C3B1}
\mathcal{C}_{3}(u)\mathcal{B}_{1}(v) &=&  
  c_{21}(u,v)\mathcal{B}_{1}(v)\mathcal{C}_{1}(u) + c_{22}(u,v)\mathcal{B}_{1}(v)\mathcal{C}_{3}(u) + 
   c_{23}(u,v)\mathcal{B}_{1}(u)\mathcal{C}_{3}(v) \nonumber\\&+& c_{24}(u,v)\mathcal{B}_{3}(u)\mathcal{C}_{3}(v) + 
   c_{25}(u,v)\mathcal{B}_{2}(v)\mathcal{C}_{2}(u) + c_{26}(u,v)\mathcal{D}_{1}(v)\mathcal{D}_{1}(u) \nonumber\\&+& 
   c_{27}(u,v)\mathcal{D}_{1}(v)\mathcal{D}_{2}(u) + c_{28}(u,v)\mathcal{D}_{1}(v)\mathcal{D}_{3}(u) + 
   c_{29}(u,v)\mathcal{D}_{1}(u)\mathcal{D}_{1}(v) \nonumber\\&+& c_{210}(u,v)\mathcal{D}_{1}(u)\mathcal{D}_{2}(v) + 
   c_{211}(u,v)\mathcal{D}_{2}(u)\mathcal{D}_{1}(v) + c_{212}(u,v)\mathcal{D}_{2}(u)\mathcal{D}_{2}(v) \nonumber\\&+& 
   c_{213}(u,v)\mathcal{D}_{3}(u)\mathcal{D}_{1}(v) + c_{214}(u,v)\mathcal{D}_{3}(u)\mathcal{D}_{2}(v),
\end{eqnarray}
\begin{eqnarray}\label{B1B1}
\mathcal{B}_{1}(u)\mathcal{B}_{1}(v) &=&  
  e_{01}(u,v)\mathcal{B}_{1}(v)\mathcal{B}_{1}(u) + e_{02}(u,v)\mathcal{B}_{2}(v)\mathcal{D}_{2}(u) + 
   e_{03}(u,v)\mathcal{B}_{2}(v)\mathcal{D}_{1}(u) \nonumber\\&+& e_{04}(u,v)\mathcal{B}_{2}(u)\mathcal{D}_{1}(v) + 
   e_{05}(u,v)\mathcal{B}_{2}(u)\mathcal{D}_{2}(v),
\end{eqnarray}
\begin{eqnarray}\label{B1B3}
\mathcal{B}_{1}(u)\mathcal{B}_{3}(v) &=&  
  d_{11}(u,v)\mathcal{B}_{3}(v)\mathcal{B}_{1}(u) + d_{12}(u,v)\mathcal{B}_{1}(v)\mathcal{B}_{1}(u) + 
   d_{13}(u,v)\mathcal{B}_{2}(v)\mathcal{D}_{1}(u) \nonumber\\&+& d_{14}(u,v)\mathcal{B}_{2}(v)\mathcal{D}_{2}(u) + 
   d_{15}(u,v)\mathcal{B}_{2}(u)\mathcal{D}_{1}(v) + d_{16}(u,v)\mathcal{B}_{2}(u)\mathcal{D}_{2}(v) \nonumber\\&+& 
   d_{17}(u,v)\mathcal{B}_{2}(u)\mathcal{D}_{3}(v),
\end{eqnarray}
\begin{eqnarray}\label{D1B2}
\mathcal{D}_{1}(u)\mathcal{B}_{2}(v) &=&  
  b_{11}(u,v)\mathcal{B}_{2}(v)\mathcal{D}_{1}(u) + b_{12}(u,v)\mathcal{B}_{2}(u)\mathcal{D}_{1}(v) + 
   b_{13}(u,v)\mathcal{B}_{2}(u)\mathcal{D}_{2}(v) \nonumber\\&+& b_{14}(u,v)\mathcal{B}_{2}(u)\mathcal{D}_{3}(v) + 
   b_{15}(u,v)\mathcal{B}_{1}(u)\mathcal{B}_{1}(v) + b_{16}(u,v)\mathcal{B}_{1}(u)\mathcal{B}_{3}(v),\nonumber\\
\end{eqnarray}
\begin{eqnarray}\label{D2B2}
\mathcal{D}_{2}(u)\mathcal{B}_{2}(v) &=&  
  b_{21}(u,v)\mathcal{B}_{2}(v)\mathcal{D}_{2}(u) + b_{22}(u,v)\mathcal{B}_{2}(u)\mathcal{D}_{1}(v) + 
   b_{23}(u,v)\mathcal{B}_{2}(u)\mathcal{D}_{2}(v) \nonumber\\&+& b_{24}(u,v)\mathcal{B}_{2}(u)\mathcal{D}_{3}(v) + 
   b_{25}(u,v)\mathcal{B}_{1}(u)\mathcal{B}_{1}(v) + b_{26}(u,v)\mathcal{B}_{1}(u)\mathcal{B}_{3}(v) \nonumber\\&+& 
   b_{27}(u,v)\mathcal{B}_{3}(u)\mathcal{B}_{1}(v) + b_{28}(u,v)\mathcal{B}_{3}(u)\mathcal{B}_{3}(v) + 
   b_{29}(u,v)\mathcal{B}_{2}(v)\mathcal{D}_{1}(u),\nonumber\\
\end{eqnarray}
\begin{eqnarray}\label{D3B2}
\mathcal{D}_{3}(u)\mathcal{B}_{2}(v) &=&  
  b_{31}(u,v)\mathcal{B}_{2}(u)\mathcal{D}_{3}(v) + b_{32}(u,v)\mathcal{B}_{2}(u)\mathcal{D}_{1}(v) + 
   b_{33}(u,v)\mathcal{B}_{2}(u)\mathcal{D}_{2}(v) \nonumber\\&+& b_{34}(u,v)\mathcal{B}_{2}(v)\mathcal{D}_{3}(u) + 
   b_{35}(u,v)\mathcal{B}_{1}(u)\mathcal{B}_{1}(v) + b_{36}(u,v)\mathcal{B}_{1}(u)\mathcal{B}_{3}(v) \nonumber\\&+& 
   b_{37}(u,v)\mathcal{B}_{3}(u)\mathcal{B}_{1}(v) + b_{38}(u,v)\mathcal{B}_{3}(u)\mathcal{B}_{3}(v) + 
   b_{39}(u,v)\mathcal{B}_{2}(v)\mathcal{D}_{1}(u) \nonumber\\&+& b_{310}(u,v)\mathcal{B}_{2}(v)\mathcal{D}_{2}(u),
\end{eqnarray}
\begin{eqnarray}\label{C1B2}
\mathcal{C}_{1}(u)\mathcal{B}_{2}(v) &=&  
  Y_{11}(u,v)\mathcal{B}_{2}(v)\mathcal{C}_{1}(u) + Y_{12}(u,v)\mathcal{B}_{2}(v)\mathcal{C}_{3}(u) + 
   Y_{13}(u,v)\mathcal{B}_{2}(u)\mathcal{C}_{1}(v) \nonumber\\&+& Y_{14}(u,v)\mathcal{B}_{2}(u)\mathcal{C}_{3}(v) + 
   Y_{15}(u,v)\mathcal{B}_{1}(v)\mathcal{D}_{1}(u) + Y_{16}(u,v)\mathcal{B}_{1}(v)\mathcal{D}_{2}(u) \nonumber\\&+& 
   Y_{17}(u,v)\mathcal{B}_{3}(v)\mathcal{D}_{1}(u) + Y_{18}(u,v)\mathcal{B}_{3}(v)\mathcal{D}_{2}(u) + 
   Y_{19}(u,v)\mathcal{B}_{1}(u)\mathcal{D}_{1}(v) \nonumber\\&+& Y_{110}(u,v)\mathcal{B}_{1}(u)\mathcal{D}_{2}(v) + 
   Y_{111}(u,v)\mathcal{B}_{1}(u)\mathcal{D}_{3}(v) + Y_{112}(u,v)\mathcal{B}_{3}(u)\mathcal{D}_{1}(v) \nonumber\\&+& 
   Y_{113}(u,v)\mathcal{B}_{3}(u)\mathcal{D}_{2}(v) + Y_{114}(u,v)\mathcal{B}_{3}(u)\mathcal{D}_{3}(v),
\end{eqnarray}
\begin{eqnarray}\label{C2B2}
\mathcal{C}_{2}(u)\mathcal{B}_{2}(v) &=&  
  c_{31}(u,v)\mathcal{D}_{2}(u)\mathcal{D}_{3}(v) + c_{32}(u,v)\mathcal{B}_{2}(u)\mathcal{C}_{2}(v) + 
   c_{33}(u,v)\mathcal{D}_{3}(u)\mathcal{D}_{3}(v) \nonumber\\&+& c_{34}(u,v)\mathcal{D}_{2}(u)\mathcal{D}_{2}(v) + 
   c_{35}(u,v)\mathcal{B}_{3}(u)\mathcal{C}_{1}(v) + c_{36}(u,v)\mathcal{D}_{2}(v)\mathcal{D}_{3}(u) \nonumber\\&+& 
   c_{37}(u,v)\mathcal{D}_{1}(u)\mathcal{D}_{2}(v) + c_{38}(u,v)\mathcal{D}_{1}(v)\mathcal{D}_{3}(u) + 
   c_{39}(u,v)\mathcal{D}_{1}(v)\mathcal{D}_{2}(u) \nonumber\\&+& c_{310}(u,v)\mathcal{B}_{3}(v)\mathcal{C}_{1}(u) + 
   c_{311}(u,v)\mathcal{B}_{1}(u)\mathcal{C}_{1}(v) + c_{312}(u,v)\mathcal{D}_{3}(u)\mathcal{D}_{1}(v) \nonumber\\&+& 
   c_{313}(u,v)\mathcal{D}_{3}(v)\mathcal{D}_{1}(u) + c_{314}(u,v)\mathcal{D}_{2}(v)\mathcal{D}_{2}(u) + 
   c_{315}(u,v)\mathcal{D}_{3}(u)\mathcal{D}_{2}(v) \nonumber\\&+& c_{316}(u,v)\mathcal{D}_{1}(u)\mathcal{D}_{3}(v) + 
   c_{317}(u,v)\mathcal{B}_{2}(v)\mathcal{C}_{2}(u) + c_{318}(u,v)\mathcal{D}_{1}(v)\mathcal{D}_{1}(u) \nonumber\\&+& 
   c_{319}(u,v)\mathcal{D}_{1}(u)\mathcal{D}_{1}(v) + c_{320}(u,v)\mathcal{D}_{2}(u)\mathcal{D}_{1}(v) + 
   c_{321}(u,v)\mathcal{B}_{3}(v)\mathcal{C}_{3}(u) \nonumber\\&+& c_{322}(u,v)\mathcal{B}_{1}(v)\mathcal{C}_{1}(u) + 
   c_{323}(u,v)\mathcal{D}_{2}(v)\mathcal{D}_{1}(u) + c_{324}(u,v)\mathcal{B}_{1}(v)\mathcal{C}_{3}(u) \nonumber\\&+& 
   c_{325}(u,v)\mathcal{B}_{3}(u)\mathcal{C}_{3}(v),
\end{eqnarray}
\begin{eqnarray}\label{C3B2}
\mathcal{C}_{3}(u)\mathcal{B}_{2}(v) &=&  
  Y_{21}(u,v)\mathcal{B}_{2}(v)\mathcal{C}_{3}(u) + Y_{22}(u,v)\mathcal{B}_{2}(v)\mathcal{C}_{1}(u) + 
   Y_{23}(u,v)\mathcal{B}_{2}(u)\mathcal{C}_{1}(v) \nonumber\\&+& Y_{24}(u,v)\mathcal{B}_{2}(u)\mathcal{C}_{3}(v) + 
   Y_{25}(u,v)\mathcal{B}_{1}(v)\mathcal{D}_{1}(u) + Y_{26}(u,v)\mathcal{B}_{1}(v)\mathcal{D}_{2}(u) \nonumber\\&+& 
   Y_{27}(u,v)\mathcal{B}_{1}(v)\mathcal{D}_{3}(u) + Y_{28}(u,v)\mathcal{B}_{3}(v)\mathcal{D}_{1}(u) + 
   Y_{29}(u,v)\mathcal{B}_{3}(v)\mathcal{D}_{2}(u) \nonumber\\&+& Y_{210}(u,v)\mathcal{B}_{3}(v)\mathcal{D}_{3}(u) + 
   Y_{211}(u,v)\mathcal{B}_{1}(u)\mathcal{D}_{1}(v) + Y_{212}(u,v)\mathcal{B}_{1}(u)\mathcal{D}_{2}(v) \nonumber\\&+& 
   Y_{213}(u,v)\mathcal{B}_{1}(u)\mathcal{D}_{3}(v) + Y_{214}(u,v)\mathcal{B}_{3}(u)\mathcal{D}_{1}(v) + 
   Y_{215}(u,v)\mathcal{B}_{3}(u)\mathcal{D}_{2}(v) \nonumber\\&+& Y_{216}(u,v)\mathcal{B}_{3}(u)\mathcal{D}_{3}(v),
\end{eqnarray}
\begin{eqnarray}\label{D1B3}
\mathcal{D}_{1}(u)\mathcal{B}_{3}(v) &=&  
  x_{11}(u,v)\mathcal{B}_{3}(v)\mathcal{D}_{1}(u) + x_{12}(u,v)\mathcal{B}_{1}(v)\mathcal{D}_{1}(u) + 
   x_{13}(u,v)\mathcal{B}_{1}(u)\mathcal{D}_{1}(v) \nonumber\\&+& x_{14}(u,v)\mathcal{B}_{1}(u)\mathcal{D}_{2}(v) + 
   x_{15}(u,v)\mathcal{B}_{1}(u)\mathcal{D}_{3}(v) + x_{16}(u,v)\mathcal{B}_{2}(u)\mathcal{C}_{1}(v) \nonumber\\&+& 
   x_{17}(u,v)\mathcal{B}_{2}(u)\mathcal{C}_{3}(v) + x_{18}(u,v)\mathcal{B}_{2}(v)\mathcal{C}_{1}(u),
\end{eqnarray}
\begin{eqnarray}\label{D2B3}
\mathcal{D}_{2}(u)\mathcal{B}_{3}(v) &=&  
  x_{21}(u,v)\mathcal{B}_{3}(v)\mathcal{D}_{2}(u) + x_{22}(u,v)\mathcal{B}_{1}(v)\mathcal{D}_{2}(u) + 
   x_{23}(u,v)\mathcal{B}_{1}(u)\mathcal{D}_{1}(v) \nonumber\\&+& x_{24}(u,v)\mathcal{B}_{1}(u)\mathcal{D}_{2}(v) + 
   x_{25}(u,v)\mathcal{B}_{1}(u)\mathcal{D}_{3}(v) + x_{26}(u,v)\mathcal{B}_{3}(u)\mathcal{D}_{1}(v) \nonumber\\&+& 
   x_{27}(u,v)\mathcal{B}_{3}(u)\mathcal{D}_{2}(v) + x_{28}(u,v)\mathcal{B}_{3}(u)\mathcal{D}_{3}(v) + 
   x_{29}(u,v)\mathcal{B}_{2}(u)\mathcal{C}_{1}(v) \nonumber\\&+& x_{210}(u,v)\mathcal{B}_{2}(u)\mathcal{C}_{3}(v) + 
   x_{211}(u,v)\mathcal{B}_{2}(v)\mathcal{C}_{1}(u) + x_{212}(u,v)\mathcal{B}_{2}(v)\mathcal{C}_{3}(u),\nonumber\\
\end{eqnarray}
\begin{eqnarray}\label{D3B3}
\mathcal{D}_{3}(u)\mathcal{B}_{3}(v) &=&  
  x_{31}(u,v)\mathcal{B}_{3}(v)\mathcal{D}_{3}(u) + x_{32}(u,v)\mathcal{B}_{1}(v)\mathcal{D}_{3}(u) + 
   x_{33}(u,v)\mathcal{B}_{1}(u)\mathcal{D}_{1}(v) \nonumber\\&+& x_{34}(u,v)\mathcal{B}_{1}(u)\mathcal{D}_{2}(v) + 
   x_{35}(u,v)\mathcal{B}_{1}(u)\mathcal{D}_{3}(v) + x_{36}(u,v)\mathcal{B}_{3}(u)\mathcal{D}_{1}(v) \nonumber\\&+& 
   x_{37}(u,v)\mathcal{B}_{3}(u)\mathcal{D}_{2}(v) + x_{38}(u,v)\mathcal{B}_{3}(u)\mathcal{D}_{3}(v) + 
   x_{39}(u,v)\mathcal{B}_{2}(u)\mathcal{C}_{1}(v) \nonumber\\&+& x_{310}(u,v)\mathcal{B}_{2}(u)\mathcal{C}_{3}(v) + 
   x_{311}(u,v)\mathcal{B}_{2}(v)\mathcal{C}_{1}(u) + x_{312}(u,v)\mathcal{B}_{2}(v)\mathcal{C}_{3}(u) \nonumber\\&+& 
   x_{313}(u,v)\mathcal{B}_{1}(v)\mathcal{D}_{1}(u) + x_{314}(u,v)\mathcal{B}_{3}(v)\mathcal{D}_{1}(u),
\end{eqnarray}
\begin{eqnarray}\label{B2B1}
\mathcal{B}_{2}(u)\mathcal{B}_{1}(v) &=&  
  e_{11}(u,v)\mathcal{B}_{1}(v)\mathcal{B}_{2}(u) + e_{12}(u,v)\mathcal{B}_{2}(v)\mathcal{B}_{1}(u) + 
   e_{13}(u,v)\mathcal{B}_{2}(v)\mathcal{B}_{3}(u),\nonumber\\
\end{eqnarray}
\begin{eqnarray}\label{B1B2}
\mathcal{B}_{1}(u)\mathcal{B}_{2}(v) &=&  
  e_{21}(u,v)\mathcal{B}_{2}(v)\mathcal{B}_{1}(u) + e_{22}(u,v)\mathcal{B}_{2}(v)\mathcal{B}_{3}(u) + 
   e_{23}(u,v)\mathcal{B}_{1}(v)\mathcal{B}_{2}(u) \nonumber\\&+& e_{24}(u,v)\mathcal{B}_{3}(v)\mathcal{B}_{2}(u),
\end{eqnarray}
\begin{eqnarray}\label{B2B3}
\mathcal{B}_{2}(u)\mathcal{B}_{3}(v) &=&  
  e_{31}(u,v)\mathcal{B}_{3}(v)\mathcal{B}_{2}(u) + e_{32}(u,v)\mathcal{B}_{1}(v)\mathcal{B}_{2}(u) + 
   e_{33}(u,v)\mathcal{B}_{2}(v)\mathcal{B}_{1}(u) \nonumber\\&+& e_{34}(u,v)\mathcal{B}_{2}(v)\mathcal{B}_{3}(u),
\end{eqnarray}
\begin{eqnarray}\label{C1B3}
\mathcal{C}_{1}(u)\mathcal{B}_{3}(v) &=&  
  C_{21}(u,v)\mathcal{B}_{1}(v)\mathcal{C}_{1}(u) + C_{22}(u,v)\mathcal{B}_{3}(v)\mathcal{C}_{1}(u) + 
   C_{23}(u,v)\mathcal{B}_{3}(u)\mathcal{C}_{1}(v) \nonumber\\&+& C_{24}(u,v)\mathcal{B}_{3}(u)\mathcal{C}_{3}(v) + 
   C_{25}(u,v)\mathcal{B}_{2}(v)\mathcal{C}_{2}(u) + C_{26}(u,v)\mathcal{D}_{1}(v)\mathcal{D}_{1}(u) \nonumber\\&+& 
   C_{27}(u,v)\mathcal{D}_{2}(v)\mathcal{D}_{1}(u) + C_{28}(u,v)\mathcal{D}_{3}(v)\mathcal{D}_{1}(u) + 
   C_{29}(u,v)\mathcal{D}_{1}(u)\mathcal{D}_{1}(v) \nonumber\\&+& C_{210}(u,v)\mathcal{D}_{2}(u)\mathcal{D}_{1}(v) + 
   C_{211}(u,v)\mathcal{D}_{1}(u)\mathcal{D}_{2}(v) + C_{212}(u,v)\mathcal{D}_{2}(u)\mathcal{D}_{2}(v) \nonumber\\&+& 
   C_{213}(u,v)\mathcal{D}_{1}(u)\mathcal{D}_{3}(v) + C_{214}(u,v)\mathcal{D}_{2}(u)\mathcal{D}_{3}(v),
\end{eqnarray}
\begin{eqnarray}\label{C3B3}
\mathcal{C}_{3}(u)\mathcal{B}_{3}(v) &=&  
  c_{41}(u,v)\mathcal{D}_{1}(v)\mathcal{D}_{2}(u) + c_{42}(u,v)\mathcal{D}_{1}(u)\mathcal{D}_{2}(v) + 
   c_{43}(u,v)\mathcal{D}_{2}(u)\mathcal{D}_{1}(v) \nonumber\\&+& c_{44}(u,v)\mathcal{B}_{1}(u)\mathcal{C}_{1}(v) + 
   c_{45}(u,v)\mathcal{D}_{2}(v)\mathcal{D}_{3}(u) + c_{46}(u,v)\mathcal{B}_{3}(v)\mathcal{C}_{1}(u) \nonumber\\&+& 
   c_{47}(u,v)\mathcal{B}_{2}(v)\mathcal{C}_{2}(u) + c_{48}(u,v)\mathcal{D}_{3}(u)\mathcal{D}_{1}(v) + 
   c_{49}(u,v)\mathcal{D}_{2}(v)\mathcal{D}_{1}(u) \nonumber\\&+& c_{410}(u,v)\mathcal{D}_{3}(u)\mathcal{D}_{2}(v) + 
   c_{411}(u,v)\mathcal{B}_{1}(v)\mathcal{C}_{3}(u) + c_{412}(u,v)\mathcal{D}_{2}(v)\mathcal{D}_{2}(u) \nonumber\\&+& 
   c_{413}(u,v)\mathcal{B}_{1}(v)\mathcal{C}_{1}(u) + c_{414}(u,v)\mathcal{D}_{1}(u)\mathcal{D}_{3}(v) + 
   c_{415}(u,v)\mathcal{D}_{1}(v)\mathcal{D}_{3}(u) \nonumber\\&+& c_{416}(u,v)\mathcal{D}_{2}(u)\mathcal{D}_{2}(v) + 
   c_{417}(u,v)\mathcal{D}_{1}(v)\mathcal{D}_{1}(u) + c_{418}(u,v)\mathcal{D}_{1}(u)\mathcal{D}_{1}(v) \nonumber\\&+& 
   c_{419}(u,v)\mathcal{B}_{2}(u)\mathcal{C}_{2}(v) + c_{420}(u,v)\mathcal{D}_{3}(v)\mathcal{D}_{1}(u) + 
   c_{421}(u,v)\mathcal{B}_{3}(u)\mathcal{C}_{3}(v) \nonumber\\&+& c_{422}(u,v)\mathcal{B}_{3}(v)\mathcal{C}_{3}(u) + 
   c_{423}(u,v)\mathcal{B}_{3}(u)\mathcal{C}_{1}(v) + c_{424}(u,v)\mathcal{D}_{3}(u)\mathcal{D}_{3}(v) \nonumber\\&+& 
   c_{425}(u,v)\mathcal{D}_{2}(u)\mathcal{D}_{3}(v),
\end{eqnarray}
\begin{eqnarray}\label{B3B1}
\mathcal{B}_{3}(u)\mathcal{B}_{1}(v) &=&  
  d_{21}(u,v)\mathcal{B}_{1}(v)\mathcal{B}_{3}(u) + d_{22}(u,v)\mathcal{B}_{1}(v)\mathcal{B}_{1}(u) + 
   d_{23}(u,v)\mathcal{B}_{2}(v)\mathcal{D}_{1}(u) \nonumber\\&+& d_{24}(u,v)\mathcal{B}_{2}(v)\mathcal{D}_{2}(u) + 
   d_{25}(u,v)\mathcal{B}_{2}(v)\mathcal{D}_{3}(u) + d_{26}(u,v)\mathcal{B}_{2}(u)\mathcal{D}_{1}(v) \nonumber\\&+&
   d_{27}(u,v)\mathcal{B}_{2}(u)\mathcal{D}_{2}(v),
\end{eqnarray}
\begin{eqnarray}\label{B3B3}
\mathcal{B}_{3}(u)\mathcal{B}_{3}(v) &=&
f_{01}(u,v)\mathcal{B}_{3}(v)\mathcal{B}_{3}(u)+
f_{02}(u,v)\mathcal{B}_{1}(u)\mathcal{B}_{1}(v)+
f_{03}(u,v)
   \mathcal{B}_{1}(u)\mathcal{B}_{3}(v)\nonumber\\&+&
   f_{04}(u,v)
   \mathcal{B}_{3}(u)\mathcal{B}_{1}(v)+f_{05}(u,v)
   \mathcal{B}_{2}(u)\mathcal{D}_{1}(v)+f_{06}(u,v)
   \mathcal{B}_{2}(v)\mathcal{D}_{1}(u)\nonumber\\&+&f_{07}(u,v)
   \mathcal{B}_{2}(u)\mathcal{D}_{2}(v)+f_{08}(u,v)
   \mathcal{B}_{2}(v)\mathcal{D}_{2}(u)+
   f_{09}(u,v)
   \mathcal{B}_{2}(u)\mathcal{D}_{3}(v)\nonumber\\&+&
   f_{010}(u,v)
   \mathcal{B}_{2}(v)\mathcal{D}_{3}(u).
\end{eqnarray}
Here we write explicitly only the amplitudes which appear
in the final solution (subsection 3.4). For more details, including explicit
expressions for the coefficients of the commutation relations, see for instance \cite{KurakLima}.
For the ZF model we have,
\begin{equation}
 a_{11}(u,v)=\frac{\sinh (u+v) \sinh (u-v-\eta )}{\sinh (u-v)
   \sinh (u+v+\eta )},
\end{equation}
\begin{equation}
 a_{21}(u,v)=\frac{\sinh (u+v) \sinh (u-v-\eta ) \sinh
   \left(u-v+\frac{\eta }{2}\right) \sinh
   \left(u+v+\frac{3 \eta }{2}\right)}{\sinh
   (u-v) \sinh \left(u-v-\frac{\eta }{2}\right)
   \sinh \left(u+v+\frac{\eta }{2}\right) \sinh
   (u+v+\eta )},
\end{equation}
\begin{equation}
 a_{31}(u,v)=\frac{\sinh \left(u-v+\frac{\eta }{2}\right)
   \sinh \left(u+v+\frac{3 \eta
   }{2}\right)}{\sinh \left(u-v-\frac{\eta
   }{2}\right) \sinh \left(u+v+\frac{\eta
   }{2}\right)},
\end{equation}
\begin{equation}
 e_{01}(u,v)=\frac{\sinh (u-v-\eta ) \sinh
   \left(u-v+\frac{\eta }{2}\right)}{\sinh
   \left(u-v-\frac{\eta }{2}\right) \sinh
   (u-v+\eta )},
\end{equation}
\begin{equation}
 e_{04}(u,v)=\frac{\sinh (2v) \sinh \left(\eta
   \right)}{\sinh \left(u-v-\frac{\eta
   }{2}\right) \sinh \left(2v+\eta\right)},~~ e_{05}(u,v)=-\frac{\sinh \left(\eta\right)
   }{\sinh
   \left(u+v+\frac{\eta }{2}\right)},
\end{equation}
and, for the IK solution, we have,
\begin{equation}
 a_{11}(u,v)=\frac{\sinh (u+v) \sinh (u-v-\eta )}{\sinh (u-v)
   \sinh (u+v+\eta )},
\end{equation}
\begin{equation}
 a_{21}(u,v)=\frac{\sinh
   (u-v+\eta ) \sinh (u+v+2 \eta )\cosh \left(u-v-\frac{\eta }{2}\right)
   \cosh \left(u+v+\frac{\eta }{2}\right) }
   {\sinh
   (u-v) \sinh (u+v+\eta )\cosh
   \left(u-v+\frac{\eta }{2}\right) \cosh
   \left(u+v+\frac{3 \eta }{2}\right)},
\end{equation}
\begin{equation}
 a_{31}(u,v)=\frac{\cosh \left(u-v+\frac{3 \eta }{2}\right)
   \cosh \left(u+v+\frac{5 \eta
   }{2}\right)}{\cosh \left(u-v+\frac{\eta
   }{2}\right) \cosh \left(u+v+\frac{3 \eta
   }{2}\right)},
\end{equation}
\begin{equation}
 e_{01}(u,v)=\frac{\cosh \left(u-v-\frac{\eta
   }{2}\right)}{\cosh \left(u-v+\frac{\eta
   }{2}\right)},
\end{equation}
\begin{equation}
 e_{04}(u,v)=\frac{e^{\eta } \sinh (2 v) \sinh (\eta )}{\cosh
   \left(u-v+\frac{\eta }{2}\right) \sinh (2
   v+\eta )},~~e_{05}(u,v)=-\frac{e^{\eta } \sinh (\eta )}{\cosh
   \left(u+v+\frac{3 \eta }{2}\right)}.
\end{equation}

\section{Coefficients of the expansions $t_{d}(u)\Psi_n$ and $t_{u}(u)\Psi_n$}
\label{coetdtu}
\setcounter{equation}{0}
\renewcommand{\theequation}{B.\arabic{equation}}

The functions entering equation (\ref{tdpsin})
are given by,
\begin{equation}\label{lam}
 \Lambda_n(u,u_1,\ldots,u_n)=\sum_{\alpha=1}^3\omega_\alpha(u)\Delta_{\alpha}(u)
 \prod_{j=1}^{n}a_{\alpha 1}(u,u_{j}),
\end{equation}
\begin{eqnarray}
\mathcal{F}_j^{(n)}(u,u_1,\ldots,u_n)&=& \left\{\prod_{p=1,p<j}^n \Omega(u_j,u_p) \right\}
 \left\{ 
 \sum_{\alpha=1}^2\Delta_\alpha(u_j) Q_\alpha^{\mathcal{F}}(u,u_j)\prod_{m=1,m\neq j}^n
 a_{\alpha 1}(u_j,u_m)\right\},\nonumber\\
\end{eqnarray}
\begin{eqnarray}
\mathcal{G}_j^{(n)}(u,u_1,\ldots,u_n)&=& \left\{\prod_{p=1,p<j}^n \Omega(u_j,u_p) \right\}
 \left\{ 
 \sum_{\alpha=1}^2\Delta_\alpha(u_j) Q_\alpha^{\mathcal{G}}(u,u_j)\prod_{m=1,m\neq j}^n
 a_{\alpha 1}(u_j,u_m)\right\},\nonumber\\
\end{eqnarray}
and
\begin{eqnarray}
\mathcal{H}_{jk}^{(n)}(u,u_1,\ldots,u_n)&=&
\left\{\prod_{p=1,p<j}^n \Omega(u_j,u_p) \right\}
\left\{\prod_{q=1,q<k,q\neq j}^n \Omega(u_k,u_q) \right\}
\nonumber\\
&\times&
\left\{
\sum_{\alpha,\beta=1}^2
\Delta_\alpha(u_j)\Delta_\beta(u_k)Q_{\alpha\beta}^\mathcal{H}(u,u_j,u_k)
\prod_{\substack{m=1\\m\neq j,k}}^n a_{\alpha 1}(u_j,u_m) a_{\beta 1}(u_k,u_m)\right\}\nonumber\\
\end{eqnarray}
with the auxiliary functions $Q$ defined by, 
\begin{equation}
 Q_1^{\mathcal{F}}(u,u_j)=\sum_{q=1}^3 \omega_q(u) a_{q2}(u,u_j),~~~
 Q_2^{\mathcal{F}}(u,u_j)=\sum_{q=1}^3 \omega_q(u) a_{q3}(u,u_j),
\end{equation}
\begin{equation}
 Q_1^{\mathcal{G}}(u,u_j)=\sum_{q=2}^3 \omega_q(u) a_{q4}(u,u_j),~~~
 Q_2^{\mathcal{G}}(u,u_j)=\sum_{q=2}^3 \omega_q(u) a_{q5}(u,u_j),
\end{equation}
\begin{eqnarray}
 Q_{11}^\mathcal{H}(u,u_j,u_k)&=&\omega_{1}(u) \left\{a_{11}(u,u_{j})
   a_{12}(u,u_{k})
   e_{03}(u_{j},u)+a_{14}(u,u_{j})
   \left[c_{16}(u_{j},u_{k}) \right. \right. \nonumber\\&+& \left. c_{18}(u_{j}
   ,u_{k})\right] + a_{15}(u,u_{j})
   \left[c_{26}(u_{j},u_{k})+c_{29}(u_{j}
   ,u_{k})\right]\nonumber\\&-& \left. b_{12}(u,u_{j})
   e_{04}(u_{j},u_{k})\right\}\nonumber\\&+&\omega_{2}(u)
   \left\{a_{21}(u,u_{j})\left[ a_{22}(u,u_{k})
   e_{03}(u_{j},u)+
   a_{24}(u,u_{k})
   d_{13}(u_{j},u)\right]\right. \nonumber\\&+&a_{26}(u,u_{j})
   \left[c_{16}(u_{j},u_{k})+c_{18}(u_{j}
   ,u_{k})\right]+a_{27}(u,u_{j})
   \left[c_{26}(u_{j},u_{k})\right.\nonumber\\&+&\left. c_{29}(u_{j}
   ,u_{k})\right]-b_{22}(u,u_{j})
   e_{04}(u_{j},u_{k})\left.\right\}\nonumber\\&+&\omega_{3}(u)
   \left\{a_{31}(u,u_{j})\left[a_{32}(u,u_{k})
   e_{03}(u_{j},u)+
   a_{34}(u,u_{k})
   d_{13}(u_{j},u)\right]\right. \nonumber\\&+&a_{36}(u,u_{j})
   \left[c_{16}(u_{j},u_{k})+c_{18}(u_{j}
   ,u_{k})\right]+a_{37}(u,u_{j})
   \left[c_{26}(u_{j},u_{k})\right.\nonumber\\&+&\left. c_{29}(u_{j}
   ,u_{k})\right]- \left.b_{32}(u,u_{j})
   e_{04}(u_{j},u_{k})\right\},
\end{eqnarray}
\begin{eqnarray}
 Q_{12}^\mathcal{H}(u,u_j,u_k)&=&\omega_{1}(u) \left\{a_{11}(u,u_{j})
   a_{13}(u,u_{k})
   e_{03}(u_{j},u) \right. +a_{14}(u,u_{j})
  c_{19}(u_{j},u_{k})\nonumber\\&+&a_{15}(u,u_{j})
  c_{210}(u_{j},u_{k})- \left. b_{12}(u,u_{j}) e_{05}(u_{j},u_{k}) \right\}\nonumber\\&+&
   \omega_{2}(u)
   \left\{
   a_{21}(u,u_{j})\right. \left[a_{23}(u,u_{k})
   e_{03}(u_{j},u)+
   a_{25}(u,u_{k})
   d_{13}(u_{j},u)\right]\nonumber\\&+&a_{26}(u,u_{j})
  c_{19}(u_{j},u_{k})+a_{27}(u,u_{j})
  c_{210}(u_{j},u_{k}) - \left. b_{22}(u,u_{j}) e_{05}(u_{j},u_{k})\right\}\nonumber\\&+&
   \omega_{3}(u)
   \left\{a_{31}(u,u_{j})\left[ a_{33}(u,u_{k})
   e_{03}(u_{j},u)+
   a_{35}(u,u_{k})
   d_{13}(u_{j},u)\right]\right.\nonumber\\&+&a_{36}(u,u_{j})
  c_{19}(u_{j},u_{k})+a_{37}(u,u_{j})
  c_{210}(u_{j},u_{k})-\left. b_{32}(u,u_{j}) e_{05}(u_{j},u_{k})\right\},\nonumber\\
\end{eqnarray}
\begin{eqnarray}
Q_{21}^\mathcal{H}(u,u_j,u_k)&=&\omega_{1}(u) \left\{a_{11}(u,u_{j})
   a_{12}(u,u_{k})
   e_{02}(u_{j},u)+ \right. a_{14}(u,u_{j})
   \left[c_{17}(u_{j},u_{k})\right. \nonumber\\&+&\left. c_{110}(u_{j},u_{k})\right]+a_{15}(u,u_{j})
   \left[c_{27}(u_{j},u_{k})+c_{211}(u_{j},u_{k})\right]\nonumber\\&-& \left. b_{13}(u,u_{j})
   e_{04}(u_{j},u_{k})\right\}\nonumber\\&+&\omega_{2}(u)
   \left\{a_{21}(u,u_{j})\right.\left[a_{22}(u,u_{k})
   e_{02}(u_{j},u)+
   a_{24}(u,u_{k})
   d_{14}(u_{j},u)\right]\nonumber\\&+&a_{26}(u,u_{j})
   \left[c_{17}(u_{j},u_{k})+c_{110}(u_{j},u_{k})\right]+a_{27}(u,u_{j})
   \left[c_{27}(u_{j},u_{k})\right. \nonumber\\&+&\left. c_{211}(u_{j},u_{k})\right]- \left. b_{23}(u,u_{j})
   e_{04}(u_{j},u_{k}) \right\}\nonumber\\&+&\omega_{3}(u)
   \left\{a_{31}(u,u_{j})\left[a_{32}(u,u_{k})
   e_{02}(u_{j},u)+
   a_{34}(u,u_{k})
   d_{14}(u_{j},u)\right] \right. \nonumber\\&+&a_{36}(u,u_{j})
   \left[c_{17}(u_{j},u_{k})+c_{110}(u_{j},u_{k})\right]+a_{37}(u,u_{j})
   \left[c_{27}(u_{j},u_{k})\right.\nonumber\\&+&\left. c_{211}(u_{j},u_{k})\right]- \left. b_{33}(u,u_{j})
   e_{04}(u_{j},u_{k})\right\},
\end{eqnarray}
\begin{eqnarray}
 Q_{22}^\mathcal{H}(u,u_j,u_k)&=&\omega_{1}(u)\left\{a_{11}(u,u_{j})
   a_{13}(u,u_{k})
   e_{02}(u_{j},u) \right. +a_{14}(u,u_{j})
  c_{111}(u_{j},u_{k})\nonumber\\&+&a_{15}(u,u_{j})
  c_{212}(u_{j},u_{k})- \left. b_{13}(u,u_{j}) e_{05}(u_{j},u_{k})\right\} \nonumber\\&+&\omega_{2}(u)
   \left\{a_{21}(u,u_{j})\left[ a_{23}(u,u_{k})
   e_{02}(u_{j},u) +
   a_{25}(u,u_{k})
   d_{14}(u_{j},u)\right]\right. \nonumber\\&+&a_{26}(u,u_{j})
  c_{111}(u_{j},u_{k})+a_{27}(u,u_{j})
  c_{212}(u_{j},u_{k})-\left. b_{23}(u,u_{j}) e_{05}(u_{j},u_{k})\right\}\nonumber\\&+&\omega_{3}(u)
   \left\{a_{31}(u,u_{j})\left[a_{33}(u,u_{k})
   e_{02}(u_{j},u)+
   a_{35}(u,u_{k})
   d_{14}(u_{j},u)\right]\right. \nonumber\\&+&a_{36}(u,u_{j})
  c_{111}(u_{j},u_{k})+a_{37}(u,u_{j})
  c_{212}(u_{j},u_{k})-\left. b_{33}(u,u_{j}) e_{05}(u_{j},u_{k})\right\}.\nonumber\\
\end{eqnarray}
For the expansion (\ref{tupsin}) we have,
\begin{eqnarray}
 \mathcal{T}_j^{(n)}(u,u_1,\ldots,u_n)&=&\left\{\prod_{p=1,p<j}^n \Omega(u_j,u_p) \right\} \nonumber\\
 &\times&
 \left\{\sum_{\alpha=1}^3\sum_{\beta=1}^2
 \Delta_\alpha(u)\Delta_\beta(u_j)Q_{\alpha\beta}^{\mathcal{T}}(u,u_j)
 \prod_{\substack{m=1\\m\neq j}}^{n}a_{\alpha1}(u,u_{m})a_{\beta1}(u_{j},u_{m})\right\},
 \nonumber\\
\end{eqnarray}
\begin{eqnarray}
&&\mathcal{U}_{jk}^{(n)}(u,u_1,\ldots,u_n)=
\left\{\prod_{p=1,p<j}^n \Omega(u_j,u_p) \right\}
\left\{\prod_{q=1,q<k,q\neq j}^n \Omega(u_k,u_q) \right\}
\nonumber\\
&\times&
\left\{\sum_{\alpha=1}^3\sum_{\substack{\beta=1\\\gamma=1}}^2
\Delta_\alpha(u)\Delta_\beta(u_j)\Delta_\gamma(u_k)Q_{\alpha\beta\gamma}^{\mathcal{U}}(u,u_j,u_k)
\prod_{\substack{m=1\\m\neq j,k}}^n a_{\alpha1}(u,u_m)a_{\beta1}(u_j,u_m) a_{\gamma1}(u_k,u_m)\right\}\nonumber\\
\end{eqnarray}
\begin{eqnarray}
\mathcal{V}_{jk}^{(n)}(u,u_1,\ldots,u_n)&=&
\left\{\prod_{p=1,p<j}^n \Omega(u_j,u_p) \right\}
\left\{\prod_{q=1,q<k,q\neq j}^n \Omega(u_k,u_q) \right\}
\nonumber\\
&\times&
\left\{
\sum_{\alpha,\beta=1}^2
\Delta_\alpha(u_j)\Delta_\beta(u_k)Q_{\alpha\beta}^\mathcal{V}(u,u_j,u_k)
\prod_{\substack{m=1\\m\neq j,k}}^n a_{\alpha 1}(u_j,u_m) a_{\beta 1}(u_k,u_m)\right\}\nonumber\\
\end{eqnarray}
\begin{eqnarray}
\mathcal{W}_{jk}^{(n)}(u,u_1,\ldots,u_n)&=&
\left\{\prod_{p=1,p<j}^n \Omega(u_j,u_p) \right\}
\left\{\prod_{q=1,q<k,q\neq j}^n \Omega(u_k,u_q) \right\}
\nonumber\\
&\times&
\left\{
\sum_{\alpha,\beta=1}^2
\Delta_\alpha(u_j)\Delta_\beta(u_k)Q_{\alpha\beta}^\mathcal{W}(u,u_j,u_k)
\prod_{\substack{m=1\\m\neq j,k}}^n a_{\alpha 1}(u_j,u_m) a_{\beta 1}(u_k,u_m)\right\}\nonumber\\
\end{eqnarray}
\begin{eqnarray}
&&\mathcal{X}_{jk\ell}^{(n)}(u,u_1,\ldots,u_n)=\nonumber\\&&
\left\{\prod_{p=1,p<j}^n \Omega(u_j,u_p) \right\}
\left\{\prod_{q=1,q<k,q\neq j}^n \Omega(u_k,u_q) \right\}
\left\{\prod_{r=1,r<\ell,r\neq j,k}^n \Omega(u_\ell,u_r) \right\}
\nonumber\\
&\times&
\left\{\sum_{\alpha,\beta,\gamma=1}^2\right.
\Delta_\alpha(u_j)\Delta_\beta(u_k)\Delta_\gamma(u_\ell)
Q_{\alpha\beta\gamma}^{\mathcal{X}}(u,u_j,u_k,u_\ell)\times\nonumber\\
&&
\left.\prod_{m=1,m\neq j,k,\ell}^n a_{\alpha1}(u_j,u_m) a_{\beta1}(u_k,u_m) a_{\gamma1}(u_\ell,u_m)
\right\},\nonumber\\
\end{eqnarray}
\begin{eqnarray}
&&\mathcal{Y}_{jk\ell}^{(n)}(u,u_1,\ldots,u_n)=\nonumber\\&&
\left\{\prod_{p=1,p<j}^n \Omega(u_j,u_p) \right\}
\left\{\prod_{q=1,q<k,q\neq j}^n \Omega(u_k,u_q) \right\}
\left\{\prod_{r=1,r<\ell,r\neq j,k}^n \Omega(u_\ell,u_r) \right\}
\nonumber\\
&\times&
\left\{\sum_{\alpha,\beta,\gamma=1}^2\right.
\Delta_\alpha(u_j)\Delta_\beta(u_k)\Delta_\gamma(u_\ell)
Q_{\alpha\beta\gamma}^{\mathcal{Y}}(u,u_j,u_k,u_\ell)\times\nonumber\\
&&
\left.\prod_{m=1,m\neq j,k,\ell}^n a_{\alpha1}(u_j,u_m) a_{\beta1}(u_k,u_m) a_{\gamma1}(u_\ell,u_m)
\right\},\nonumber\\
\end{eqnarray}
\begin{eqnarray}
&&\mathcal{Z}_{jk\ell}^{(n)}(u,u_1,\ldots,u_n)=\nonumber\\&&
\left\{\prod_{p=1,p<j}^n \Omega(u_j,u_p) \right\}
\left\{\prod_{q=1,q<k,q\neq j}^n \Omega(u_k,u_q) \right\}
\left\{\prod_{r=1,r<\ell,r\neq j,k}^n \Omega(u_\ell,u_r) \right\}
\nonumber\\
&\times&
\left\{\sum_{\alpha,\beta,\gamma=1}^2\right.
\Delta_\alpha(u_j)\Delta_\beta(u_k)\Delta_\gamma(u_\ell)
Q_{\alpha\beta\gamma}^{\mathcal{Z}}(u,u_j,u_k,u_\ell)\times\nonumber\\
&&
\left.\prod_{m=1,m\neq j,k,\ell}^n a_{\alpha1}(u_j,u_m) a_{\beta1}(u_k,u_m) a_{\gamma1}(u_\ell,u_m)
\right\}.\nonumber\\
\end{eqnarray}
We have the following expressions,
\begin{equation}
Q_{11}^{\mathcal{T}}(u,u_j)=
k_{12}^{+}(u)[c_{16}(u,u_j)+c_{18}(u,u_j)]+
k_{23}^{+}(u)[c_{26}(u,u_j)+c_{29}(u,u_j)],
\end{equation}
\begin{equation}
Q_{12}^{\mathcal{T}}(u,u_j)=
k_{12}^{+}(u)c_{19}(u,u_j)+
k_{23}^{+}(u)c_{210}(u,u_j),
\end{equation}
\begin{equation}
Q_{21}^{\mathcal{T}}(u,u_j)=k_{12}^{+}(u)[c_{110}(u,u_j)+c_{17}(u,u_j)]+
k_{23}^{+}(u)[c_{211}(u,u_j)+c_{27}(u,u_j)],
\end{equation}
\begin{equation}
Q_{22}^{\mathcal{T}}(u,u_j)=k_{12}^{+}(u)c_{111}(u,u_j)+
k_{23}^{+}(u)c_{212}(u,u_j),
\end{equation}
\begin{equation}
Q_{31}^{\mathcal{T}}(u,u_j)=
k_{23}^{+}(u)[c_{213}(u,u_j)+c_{28}(u,u_j)],~~Q_{32}^{\mathcal{T}}(u,u_j)=
k_{23}^{+}(u)c_{214}(u,u_j),
\end{equation}
\begin{eqnarray}
&&Q_{111}^{\mathcal{U}}(u,u_j,u_k)=\nonumber\\
&& k^{+}_{13}(u) \left[y_{15}\left(u,u_j\right)
   \left(c_{16}\left(u,u_k\right) + c_{18}\left(u,u_k\right)\right)
   +
   y_{19}\left(u,u_j\right)
   \left(c_{16}\left(u_j,u_k\right)+c_{18}\left(u_j,u_k\right)\right)
   \right.\nonumber\\&&+\left.
   y_{112}\left(u,u_j\right)
   \left(c_{26}\left(u_j,u_k\right)+c_{29}\left(u_j,u_k\right)\right)
   +
   y_{17}\left(u,u_j\right)
   \left(c_{26}\left(u,u_k\right)+c_{29}\left(u,u_k\right)\right)
   \right.\nonumber\\&&-\left.
   e_{04}\left(u_j,u_k\right)\left(c_{318}\left(u,u_j\right)+c_{319
   }\left(u,u_j\right)\right)  \right],
\end{eqnarray}
\begin{eqnarray}
&&Q_{112}^{\mathcal{U}}(u,u_j,u_k)=k^{+}_{13}(u) \left[
c_{19}\left(u,u_k\right) y_{15}\left(u,u_j\right)+y_{19}\left(u,u_j\right)
   c_{19}\left(u_j,u_k\right)\right.\nonumber\\&&+\left.y_{112}\left(u,u_j\right)
   c_{210}\left(u_j,u_k\right)+c_{210}\left(u,u_k\right)
   y_{17}\left(u,u_j\right)
   \right.\nonumber\\&&-\left.
   e_{05}\left(u_j,u_k\right)\left(c_{318}\left(u,u_j\right)+c_{319}\left(u,u_j\right)\right)
   \right],
\end{eqnarray}
\begin{eqnarray}
&&Q_{121}^{\mathcal{U}}(u,u_j,u_k)=\nonumber\\&&
k^{+}_{13}(u) \left[y_{19}\left(u,u_j\right)
   \left(c_{110}\left(u_j,u_k\right)+c_{17}\left(u_j,u_k\right)\right)
   +
   y_{16}\left(u,u_j\right)
   \left(c_{16}\left(u,u_k\right)+c_{18}\left(u,u_k\right)\right)
   \right.\nonumber\\&&+\left.
   y_{112}\left(u,u_j\right)
   \left(c_{211}\left(u_j,u_k\right)+c_{27}\left(u_j,u_k\right)\right)
   +
   y_{18}\left(u,u_j\right)
   \left(c_{26}\left(u,u_k\right)+c_{29}\left(u,u_k\right)\right)
   \right.\nonumber\\&&-\left.
   e_{04}\left(u_j,u_k\right)\left(c_{323}\left(u,u_j\right)+c_{37}
   \left(u,u_j\right)\right) \right],
\end{eqnarray}
\begin{eqnarray}
&&Q_{122}^{\mathcal{U}}(u,u_j,u_k)=k^{+}_{13}(u) \left[y_{19}\left(u,u_j\right) c_{111}\left(u_j,u_k\right)+
c_{19}\left(u,u_k\right)
   y_{16}\left(u,u_j\right)\right.\nonumber\\&&+\left.
   c_{210}\left(u,u_k\right) y_{18}\left(u,u_j\right)+y_{112}\left(u,u_j\right)
   c_{212}\left(u_j,u_k\right)
   \right.\nonumber\\&&-\left.
   e_{05}\left(u_j,u_k\right)\left(c_{323}\left(u,u_j\right)+c_{37}\left(u,u_j\right)\right)
   \right],
\end{eqnarray}
\begin{eqnarray}
&&Q_{211}^{\mathcal{U}}(u,u_j,u_k)=\nonumber\\&&
k^{+}_{13}(u) \left[y_{15}\left(u,u_j\right)
   \left(c_{110}\left(u,u_k\right)+c_{17}\left(u,u_k\right)\right)
   +
   y_{110}\left(u,u_j\right)
   \left(c_{16}\left(u_j,u_k\right)+c_{18}\left(u_j,u_k\right)\right)
   \right.\nonumber\\&&+\left.
   y_{17}\left(u,u_j\right)
   \left(c_{211}\left(u,u_k\right)+c_{27}\left(u,u_k\right)\right)
   +
   y_{113}\left(u,u_j\right)
   \left(c_{26}\left(u_j,u_k\right)+c_{29}\left(u_j,u_k\right)\right)
   \right.\nonumber\\&&-\left.
   e_{04}\left(u_j,u_k\right)\left(c_{320}\left(u,u_j\right)+c_{39}\left(u,u_j\right)\right) \right],
\end{eqnarray}
\begin{eqnarray}
&&Q_{212}^{\mathcal{U}}(u,u_j,u_k)=k^{+}_{13}(u) \left[c_{111}\left(u,u_k\right) y_{15}\left(u,u_j\right)+y_{110}\left(u,u_j\right)
   c_{19}\left(u_j,u_k\right)
   \right.\nonumber\\&&+\left.
   y_{113}\left(u,u_j\right)
   c_{210}\left(u_j,u_k\right)+c_{212}\left(u,u_k\right)
   y_{17}\left(u,u_j\right)
   \right.\nonumber\\&&-\left.
   e_{05}\left(u_j,u_k\right)\left(c_{320}\left(u,u_j\right)+c_{39}\left(u,u_j\right)\right)
   \right],
\end{eqnarray}
\begin{eqnarray}
&&Q_{221}^{\mathcal{U}}(u,u_j,u_k)=\nonumber\\&&
k^{+}_{13}(u) \left[y_{110}\left(u,u_j\right)
   \left(c_{110}\left(u_j,u_k\right)+c_{17}\left(u_j,u_k\right)\right)
   +
   y_{16}\left(u,u_j\right)
   \left(c_{110}\left(u,u_k\right)+c_{17}\left(u,u_k\right)\right)
   \right.\nonumber\\&&+\left.
   y_{113}\left(u,u_j\right)
   \left(c_{211}\left(u_j,u_k\right)+c_{27}\left(u_j,u_k\right)\right)
   +
   y_{18}\left(u,u_j\right)
   \left(c_{211}\left(u,u_k\right)+c_{27}\left(u,u_k\right)\right)
   \right.\nonumber\\&&-\left.
   e_{04}\left(u_j,u_k\right)\left(c_{314}\left(u,u_j\right)+c_{34
   }\left(u,u_j\right)\right)\right],
\end{eqnarray}
\begin{eqnarray}
&&Q_{222}^{\mathcal{U}}(u,u_j,u_k)=k^{+}_{13}(u) \left[
y_{110}\left(u,u_j\right) c_{111}\left(u_j,u_k\right)+c_{111}\left(u,u_k\right)
   y_{16}\left(u,u_j\right)\right.\nonumber\\&&+\left.
   y_{113}\left(u,u_j\right)
   c_{212}\left(u_j,u_k\right)+c_{212}\left(u,u_k\right)
   y_{18}\left(u,u_j\right)
   \right.\nonumber\\&&-\left.
   e_{05}\left(u_j,u_k\right)\left(c_{314}\left(u,u_j\right)+c_{34}\left(u,u_j\right)\right)
   \right],
\end{eqnarray}
\begin{eqnarray}
&&Q_{311}^{\mathcal{U}}(u,u_j,u_k)=\nonumber\\&&
k^{+}_{13}(u) \left[y_{111}\left(u,u_j\right)
   \left(c_{16}\left(u_j,u_k\right)+c_{18}\left(u_j,u_k\right)\right)
   +
   y_{17}\left(u,u_j\right)
   \left(c_{213}\left(u,u_k\right)+c_{28}\left(u,u_k\right)\right)
   \right.\nonumber\\&&+\left.
   y_{114}\left(u,u_j\right)
   \left(c_{26}\left(u_j,u_k\right)+c_{29}\left(u_j,u_k\right)\right)
   -
   e_{04}\left(u_j,u_k\right)\left(c_{312}\left(u,u_j\right)+c_{38}\left(u,u_j\right)\right)\right],\nonumber\\
\end{eqnarray}
\begin{eqnarray}
&&Q_{312}^{\mathcal{U}}(u,u_j,u_k)=k^{+}_{13}(u) \left[
y_{111}\left(u,u_j\right) c_{19}\left(u_j,u_k\right)+y_{114}\left(u,u_j\right)
   c_{210}\left(u_j,u_k\right)
   \right.\nonumber\\&&+\left.
   c_{214}\left(u,u_k\right)
   y_{17}\left(u,u_j\right)-e_{05}\left(u_j,u_k\right)\left(c_{312}\left(u,u_j\right)+c_{38}\left(u,u_j\right)\right)
   \right],
\end{eqnarray}
\begin{eqnarray}
&&Q_{321}^{\mathcal{U}}(u,u_j,u_k)=k^{+}_{13}(u) \left[y_{111}\left(u,u_j\right)
   \left(c_{110}\left(u_j,u_k\right)+c_{17}\left(u_j,u_k\right)\right)
   \right.\nonumber\\&&+\left.
   y_{114}\left(u,u_j\right)
   \left(c_{211}\left(u_j,u_k\right)+c_{27}\left(u_j,u_k\right)\right)
   +
   y_{18}\left(u,u_j\right)
   \left(c_{213}\left(u,u_k\right)+c_{28}\left(u,u_k\right)\right)
   \right.\nonumber\\&&-\left.
   e_{04}\left(u_j,u_k\right)\left(c_{315}\left(u,u_j\right)+c_{36
   }\left(u,u_j\right)\right) \right],
\end{eqnarray}
\begin{eqnarray}
&&Q_{322}^{\mathcal{U}}(u,u_j,u_k)=k^{+}_{13}(u) \left[
y_{111}\left(u,u_j\right) c_{111}\left(u_j,u_k\right)+y_{114}\left(u,u_j\right)
   c_{212}\left(u_j,u_k\right)
   \right.\nonumber\\&&+\left.
   c_{214}\left(u,u_k\right)
   y_{18}\left(u,u_j\right)-e_{05}\left(u_j,u_k\right)\left(c_{315}\left(u,u_j\right)+c_{36}\left(u,u_j\right)\right)
   \right],
\end{eqnarray}
\begin{eqnarray}
&&Q_{11}^{\mathcal{V}}(u,u_j,u_k)=k^{+}_{23}(u)\nonumber\\&\times& \left[
a_{11}\left(u_j,u_k\right) \left(a_{12}\left(u,u_k\right)
   c_{29}\left(u,u_j\right)+a_{22}\left(u,u_k\right) c_{211}\left(u,u_j\right)+a_{32}\left(u,u_k\right)
   c_{213}\left(u,u_j\right)\right)\right.\nonumber\\&+&\left.
   a_{11}\left(u_j,u\right) \left(a_{12}\left(u,u_k\right)
   c_{26}\left(u,u_j\right)+a_{22}\left(u,u_k\right) c_{27}\left(u,u_j\right)+a_{32}\left(u,u_k\right)
   c_{28}\left(u,u_j\right)\right)
   \right.\nonumber\\&+&\left.
   a_{12}\left(u_j,u_k\right) \left(a_{22}\left(u,u_j\right)
   c_{211}\left(u,u_j\right)+a_{32}\left(u,u_j\right)
   c_{213}\left(u,u_j\right)\right)
   \right.\nonumber\\&+&\left.
   a_{12}\left(u,u_j\right) \left(c_{29}\left(u,u_j\right)
   a_{12}\left(u_j,u_k\right)+c_{210}\left(u,u_j\right)
   a_{22}\left(u_j,u_k\right)\right)
   \right.\nonumber\\&+&\left.
   c_{214}\left(u,u_j\right) \left(a_{32}\left(u,u_j\right)
   a_{22}\left(u_j,u_k\right)+x_{33}\left(u,u_j\right)
   a_{24}\left(u_j,u_k\right)\right)
   \right.\nonumber\\&+&\left.
   a_{22}\left(u,u_j\right) c_{212}\left(u,u_j\right)
   a_{22}\left(u_j,u_k\right)
   \right.\nonumber\\&+&\left.
   x_{12}\left(u_j,u\right) \left(a_{24}\left(u,u_k\right)
   c_{27}\left(u,u_j\right)+a_{34}\left(u,u_k\right)
   c_{28}\left(u,u_j\right)\right)
   \right.\nonumber\\&+&\left.
   a_{24}\left(u_j,u_k\right) \left(c_{210}\left(u,u_j\right)
   x_{13}\left(u,u_j\right)+c_{212}\left(u,u_j\right)
   x_{23}\left(u,u_j\right)\right)
   \right.\nonumber\\&+&\left.
   c_{23}\left(u,u_j\right)
   \left(c_{26}\left(u_j,u_k\right)+c_{29}\left(u_j,u_k\right)\right)-Y_{211}\left(u,u_j\right)
   e_{04}\left(u_j,u_k\right)\right]\nonumber\\&+&
   k^{+}_{12}(u) \left[a_{11}\left(u_j,u_k\right)
   \left(a_{12}\left(u,u_k\right) c_{18}\left(u,u_j\right)+a_{22}\left(u,u_k\right)
   c_{110}\left(u,u_j\right)\right)
   \right.\nonumber\\&+&\left.
   a_{11}\left(u_j,u\right) a_{12}\left(u,u_k\right)
   c_{16}\left(u,u_j\right)
   \right.\nonumber\\&+&\left.
   c_{17}\left(u,u_j\right) \left(a_{11}\left(u_j,u\right)
   a_{22}\left(u,u_k\right)+a_{24}\left(u,u_k\right)
   x_{12}\left(u_j,u\right)\right)
   \right.\nonumber\\&+&\left.
   c_{19}\left(u,u_j\right) \left(a_{12}\left(u,u_j\right)
   a_{22}\left(u_j,u_k\right)+x_{13}\left(u,u_j\right)
   a_{24}\left(u_j,u_k\right)\right)
   \right.\nonumber\\&+&\left.
   a_{12}\left(u_j,u_k\right) \left(a_{12}\left(u,u_j\right)
   c_{18}\left(u,u_j\right)+a_{22}\left(u,u_j\right)
   c_{110}\left(u,u_j\right)\right)
   \right.\nonumber\\&+&\left.
   a_{22}\left(u,u_j\right) c_{111}\left(u,u_j\right)
   a_{22}\left(u_j,u_k\right)+c_{111}\left(u,u_j\right) x_{23}\left(u,u_j\right)
   a_{24}\left(u_j,u_k\right)
   \right.\nonumber\\&+&\left.
   c_{13}\left(u,u_j\right)
   \left(c_{26}\left(u_j,u_k\right)+c_{29}\left(u_j,u_k\right)\right)-Y_{19}\left(u,u_j\right)
   e_{04}\left(u_j,u_k\right)\right],
\end{eqnarray}
\begin{eqnarray}
&&Q_{12}^{\mathcal{V}}(u,u_j,u_k)=k^{+}_{23}(u)\nonumber\\&\times&
\left[a_{11}\left(u_j,u_k\right) \left(a_{13}\left(u,u_k\right)
   c_{29}\left(u,u_j\right)+a_{23}\left(u,u_k\right) c_{211}\left(u,u_j\right)+a_{33}\left(u,u_k\right)
   c_{213}\left(u,u_j\right)\right)
   \right.\nonumber\\&+&\left.
   a_{11}\left(u_j,u\right) \left(a_{13}\left(u,u_k\right)
   c_{26}\left(u,u_j\right)+a_{23}\left(u,u_k\right) c_{27}\left(u,u_j\right)+a_{33}\left(u,u_k\right)
   c_{28}\left(u,u_j\right)\right)
   \right.\nonumber\\&+&\left.
   a_{12}\left(u,u_j\right) \left(c_{29}\left(u,u_j\right)
   a_{13}\left(u_j,u_k\right)+c_{210}\left(u,u_j\right)
   a_{23}\left(u_j,u_k\right)\right)
   \right.\nonumber\\&+&\left.
   a_{13}\left(u_j,u_k\right) \left(a_{22}\left(u,u_j\right)
   c_{211}\left(u,u_j\right)+a_{32}\left(u,u_j\right)
   c_{213}\left(u,u_j\right)\right)
   \right.\nonumber\\&+&\left.
   a_{22}\left(u,u_j\right) c_{212}\left(u,u_j\right)
   a_{23}\left(u_j,u_k\right)
   \right.\nonumber\\&+&\left.
   c_{214}\left(u,u_j\right) \left(a_{32}\left(u,u_j\right)
   a_{23}\left(u_j,u_k\right)+x_{33}\left(u,u_j\right)
   a_{25}\left(u_j,u_k\right)\right)
   \right.\nonumber\\&+&\left.
   x_{12}\left(u_j,u\right) \left(a_{25}\left(u,u_k\right)
   c_{27}\left(u,u_j\right)+a_{35}\left(u,u_k\right)
   c_{28}\left(u,u_j\right)\right)
   \right.\nonumber\\&+&\left.
   a_{25}\left(u_j,u_k\right) \left(c_{210}\left(u,u_j\right)
   x_{13}\left(u,u_j\right)+c_{212}\left(u,u_j\right)
   x_{23}\left(u,u_j\right)\right)+c_{23}\left(u,u_j\right)
   c_{210}\left(u_j,u_k\right)
   \right.\nonumber\\&-&\left.
   Y_{211}\left(u,u_j\right) e_{05}\left(u_j,u_k\right)\right]
   \nonumber\\
   &+&k^{+}_{12}(u)
   \left[a_{11}\left(u_j,u_k\right) \left(a_{13}\left(u,u_k\right)
   c_{18}\left(u,u_j\right)+a_{23}\left(u,u_k\right)
   c_{110}\left(u,u_j\right)\right)
   \right.\nonumber\\&+&\left.
   a_{11}\left(u_j,u\right) a_{13}\left(u,u_k\right)
   c_{16}\left(u,u_j\right)
   \right.\nonumber\\&+&\left.
   c_{17}\left(u,u_j\right) \left(a_{11}\left(u_j,u\right)
   a_{23}\left(u,u_k\right)+a_{25}\left(u,u_k\right)
   x_{12}\left(u_j,u\right)\right)
   \right.\nonumber\\&+&\left.
   a_{13}\left(u_j,u_k\right) \left(a_{12}\left(u,u_j\right)
   c_{18}\left(u,u_j\right)+a_{22}\left(u,u_j\right)
   c_{110}\left(u,u_j\right)\right)
   \right.\nonumber\\&+&\left.
   c_{19}\left(u,u_j\right) \left(a_{12}\left(u,u_j\right)
   a_{23}\left(u_j,u_k\right)+x_{13}\left(u,u_j\right)
   a_{25}\left(u_j,u_k\right)\right)
   \right.\nonumber\\&+&\left.
   a_{22}\left(u,u_j\right) c_{111}\left(u,u_j\right)
   a_{23}\left(u_j,u_k\right)+c_{111}\left(u,u_j\right) x_{23}\left(u,u_j\right)
   a_{25}\left(u_j,u_k\right)
   \right.\nonumber\\&+&\left.
   c_{13}\left(u,u_j\right)
   c_{210}\left(u_j,u_k\right)-Y_{19}\left(u,u_j\right) e_{05}\left(u_j,u_k\right)\right],
\end{eqnarray}
\begin{eqnarray}
&&Q_{21}^{\mathcal{V}}(u,u_j,u_k)=k^{+}_{23}(u)\nonumber\\&\times& \left[a_{13}\left(u,u_j\right) \left(c_{29}\left(u,u_j\right)
   a_{12}\left(u_j,u_k\right)+c_{210}\left(u,u_j\right)
   a_{22}\left(u_j,u_k\right)\right)
   \right.\nonumber\\&+&\left.
   a_{12}\left(u,u_k\right) c_{210}\left(u,u_j\right)
   a_{21}\left(u_j,u_k\right)
   \right.\nonumber\\&+&\left.
   a_{12}\left(u_j,u_k\right) \left(a_{23}\left(u,u_j\right)
   c_{211}\left(u,u_j\right)+a_{33}\left(u,u_j\right)
   c_{213}\left(u,u_j\right)\right)
   \right.\nonumber\\&+&\left.
   a_{21}\left(u_j,u_k\right) \left(a_{22}\left(u,u_k\right)
   c_{212}\left(u,u_j\right)+a_{32}\left(u,u_k\right)
   c_{214}\left(u,u_j\right)\right)
   \right.\nonumber\\&+&\left.
   a_{23}\left(u,u_j\right) c_{212}\left(u,u_j\right)
   a_{22}\left(u_j,u_k\right)+a_{33}\left(u,u_j\right) c_{214}\left(u,u_j\right)
   a_{22}\left(u_j,u_k\right)
   \right.\nonumber\\&+&\left.
   c_{210}\left(u,u_j\right) x_{14}\left(u,u_j\right)
   a_{24}\left(u_j,u_k\right)+c_{212}\left(u,u_j\right) x_{24}\left(u,u_j\right)
   a_{24}\left(u_j,u_k\right)
   \right.\nonumber\\&+&\left.
   c_{214}\left(u,u_j\right) x_{34}\left(u,u_j\right)
   a_{24}\left(u_j,u_k\right)+c_{23}\left(u,u_j\right)
   \left(c_{211}\left(u_j,u_k\right)+c_{27}\left(u_j,u_k\right)\right)
   \right.\nonumber\\&-&\left.
   Y_{212}\left(u,u_j\right)
   e_{04}\left(u_j,u_k\right)\right]
   \nonumber\\&+&
   k^{+}_{12}(u) \left[a_{12}\left(u_j,u_k\right)
   \left(a_{13}\left(u,u_j\right) c_{18}\left(u,u_j\right)+a_{23}\left(u,u_j\right)
   c_{110}\left(u,u_j\right)\right)
   \right.\nonumber\\&+&\left.
   a_{21}\left(u_j,u_k\right) \left(a_{12}\left(u,u_k\right)
   c_{19}\left(u,u_j\right)+a_{22}\left(u,u_k\right)
   c_{111}\left(u,u_j\right)\right)
   \right.\nonumber\\&+&\left.
   a_{22}\left(u_j,u_k\right) \left(a_{13}\left(u,u_j\right)
   c_{19}\left(u,u_j\right)+a_{23}\left(u,u_j\right)
   c_{111}\left(u,u_j\right)\right)
   \right.\nonumber\\&+&\left.
   a_{24}\left(u_j,u_k\right) \left(c_{111}\left(u,u_j\right)
   x_{24}\left(u,u_j\right)+c_{19}\left(u,u_j\right)
   x_{14}\left(u,u_j\right)\right)
   \right.\nonumber\\&+&\left.
   c_{13}\left(u,u_j\right)
   \left(c_{211}\left(u_j,u_k\right)+c_{27}\left(u_j,u_k\right)\right)-Y_{110}\left(u,u_j\right)
   e_{04}\left(u_j,u_k\right)\right],
\end{eqnarray}
\begin{eqnarray}
&&Q_{22}^{\mathcal{V}}(u,u_j,u_k)=k^{+}_{23}(u)\nonumber\\&\times&
\left[a_{13}\left(u_j,u_k\right) \left(a_{23}\left(u,u_j\right)
   c_{211}\left(u,u_j\right)+a_{33}\left(u,u_j\right)
   c_{213}\left(u,u_j\right)\right)
   \right.\nonumber\\&+&\left.
   a_{13}\left(u,u_k\right) c_{210}\left(u,u_j\right)
   a_{21}\left(u_j,u_k\right)
   \right.\nonumber\\&+&\left.
   a_{13}\left(u,u_j\right) \left(c_{29}\left(u,u_j\right)
   a_{13}\left(u_j,u_k\right)+c_{210}\left(u,u_j\right)
   a_{23}\left(u_j,u_k\right)\right)
   \right.\nonumber\\&+&\left.
   a_{23}\left(u,u_k\right) c_{212}\left(u,u_j\right)
   a_{21}\left(u_j,u_k\right)+a_{33}\left(u,u_k\right) c_{214}\left(u,u_j\right)
   a_{21}\left(u_j,u_k\right)
   \right.\nonumber\\&+&\left.
   a_{33}\left(u,u_j\right) c_{214}\left(u,u_j\right)
   a_{23}\left(u_j,u_k\right)+a_{23}\left(u,u_j\right) c_{212}\left(u,u_j\right)
   a_{23}\left(u_j,u_k\right)
   \right.\nonumber\\&+&\left.
   c_{210}\left(u,u_j\right) x_{14}\left(u,u_j\right)
   a_{25}\left(u_j,u_k\right)+c_{212}\left(u,u_j\right) x_{24}\left(u,u_j\right)
   a_{25}\left(u_j,u_k\right)
   \right.\nonumber\\&+&\left.
   c_{214}\left(u,u_j\right) x_{34}\left(u,u_j\right)
   a_{25}\left(u_j,u_k\right)+c_{23}\left(u,u_j\right)
   c_{212}\left(u_j,u_k\right)-Y_{212}\left(u,u_j\right) e_{05}\left(u_j,u_k\right)\right]
   \nonumber\\&+&
   k^{+}_{12}(u)
   \left[a_{21}\left(u_j,u_k\right) \left(a_{13}\left(u,u_k\right)
   c_{19}\left(u,u_j\right)+a_{23}\left(u,u_k\right)
   c_{111}\left(u,u_j\right)\right)
   \right.\nonumber\\&+&\left.
   c_{19}\left(u,u_j\right) \left(a_{13}\left(u,u_j\right)
   a_{23}\left(u_j,u_k\right)+x_{14}\left(u,u_j\right)
   a_{25}\left(u_j,u_k\right)\right)
   \right.\nonumber\\&+&\left.
   a_{13}\left(u_j,u_k\right) \left(a_{13}\left(u,u_j\right)
   c_{18}\left(u,u_j\right)+a_{23}\left(u,u_j\right)
   c_{110}\left(u,u_j\right)\right)
   \right.\nonumber\\&+&\left.
   a_{23}\left(u,u_j\right) c_{111}\left(u,u_j\right)
   a_{23}\left(u_j,u_k\right)+c_{111}\left(u,u_j\right) x_{24}\left(u,u_j\right)
   a_{25}\left(u_j,u_k\right)
   \right.\nonumber\\&+&\left.
   c_{13}\left(u,u_j\right)
   c_{212}\left(u_j,u_k\right)-Y_{110}\left(u,u_j\right) e_{05}\left(u_j,u_k\right)\right],
\end{eqnarray}
\begin{eqnarray}
&&Q_{11}^{\mathcal{W}}(u,u_j,u_k)=k^{+}_{23}(u)\nonumber\\&\times&
\left[a_{11}\left(u_j,u_k\right) \left(a_{24}\left(u,u_k\right)
   c_{211}\left(u,u_j\right)+a_{34}\left(u,u_k\right)
   c_{213}\left(u,u_j\right)\right)
   \right.\nonumber\\&+&\left.
   a_{12}\left(u_j,u_k\right) \left(a_{24}\left(u,u_j\right)
   c_{211}\left(u,u_j\right)+a_{34}\left(u,u_j\right)
   c_{213}\left(u,u_j\right)\right)
   \right.\nonumber\\&+&\left.
   a_{22}\left(u_j,u_k\right) \left(a_{24}\left(u,u_j\right)
   c_{212}\left(u,u_j\right)+a_{34}\left(u,u_j\right)
   c_{214}\left(u,u_j\right)\right)
   \right.\nonumber\\&+&\left.
   x_{11}\left(u_j,u\right) \left(a_{24}\left(u,u_k\right)
   c_{27}\left(u,u_j\right)+a_{34}\left(u,u_k\right)
   c_{28}\left(u,u_j\right)\right)
   \right.\nonumber\\&+&\left.
   a_{24}\left(u_j,u_k\right) \left(c_{212}\left(u,u_j\right)
   x_{26}\left(u,u_j\right)+c_{214}\left(u,u_j\right)
   x_{36}\left(u,u_j\right)\right)
   \right.\nonumber\\&+&\left.
   c_{24}\left(u,u_j\right)
   \left(c_{26}\left(u_j,u_k\right)+c_{29}\left(u_j,u_k\right)\right)-Y_{214}\left(u,u_j\right)
   e_{04}\left(u_j,u_k\right)\right]
   \nonumber\\&+&k^{+}_{12}(u) \left[a_{24}\left(u,u_k\right)
   \left(c_{110}\left(u,u_j\right) a_{11}\left(u_j,u_k\right)+c_{17}\left(u,u_j\right)
   x_{11}\left(u_j,u\right)\right)
   \right.\nonumber\\&+&\left.
   a_{24}\left(u,u_j\right) c_{110}\left(u,u_j\right)
   a_{12}\left(u_j,u_k\right)
   \right.\nonumber\\&+&\left.
   c_{111}\left(u,u_j\right) \left(a_{24}\left(u,u_j\right)
   a_{22}\left(u_j,u_k\right)+x_{26}\left(u,u_j\right)
   a_{24}\left(u_j,u_k\right)\right)
   \right.\nonumber\\&+&\left.
   c_{14}\left(u,u_j\right)
   \left(c_{26}\left(u_j,u_k\right)+c_{29}\left(u_j,u_k\right)\right)-Y_{112}\left(u,u_j\right)
   e_{04}\left(u_j,u_k\right)\right],
\end{eqnarray}
\begin{eqnarray}
&&Q_{12}^{\mathcal{W}}(u,u_j,u_k)=k^{+}_{23}(u)\nonumber\\&\times&
\left[a_{11}\left(u_j,u_k\right) \left(a_{25}\left(u,u_k\right)
   c_{211}\left(u,u_j\right)+a_{35}\left(u,u_k\right)
   c_{213}\left(u,u_j\right)\right)
   \right.\nonumber\\&+&\left.
   a_{13}\left(u_j,u_k\right) \left(a_{24}\left(u,u_j\right)
   c_{211}\left(u,u_j\right)+a_{34}\left(u,u_j\right)
   c_{213}\left(u,u_j\right)\right)
   \right.\nonumber\\&+&\left.
   a_{23}\left(u_j,u_k\right) \left(a_{24}\left(u,u_j\right)
   c_{212}\left(u,u_j\right)+a_{34}\left(u,u_j\right)
   c_{214}\left(u,u_j\right)\right)
   \right.\nonumber\\&+&\left.
   x_{11}\left(u_j,u\right) \left(a_{25}\left(u,u_k\right)
   c_{27}\left(u,u_j\right)+a_{35}\left(u,u_k\right)
   c_{28}\left(u,u_j\right)\right)
   \right.\nonumber\\&+&\left.
   a_{25}\left(u_j,u_k\right) \left(c_{212}\left(u,u_j\right)
   x_{26}\left(u,u_j\right)+c_{214}\left(u,u_j\right)
   x_{36}\left(u,u_j\right)\right)
   \right.\nonumber\\&+&\left.
   c_{24}\left(u,u_j\right)
   c_{210}\left(u_j,u_k\right)-Y_{214}\left(u,u_j\right) e_{05}\left(u_j,u_k\right)\right]
   \nonumber\\&+&
   k^{+}_{12}(u)
   \left[a_{25}\left(u,u_k\right) \left(c_{110}\left(u,u_j\right)
   a_{11}\left(u_j,u_k\right)+c_{17}\left(u,u_j\right)
   x_{11}\left(u_j,u\right)\right)
   \right.\nonumber\\&+&\left.
   a_{24}\left(u,u_j\right) c_{110}\left(u,u_j\right)
   a_{13}\left(u_j,u_k\right)
   \right.\nonumber\\&+&\left.
   c_{111}\left(u,u_j\right) \left(a_{24}\left(u,u_j\right)
   a_{23}\left(u_j,u_k\right)+x_{26}\left(u,u_j\right)
   a_{25}\left(u_j,u_k\right)\right)
   \right.\nonumber\\&+&\left.
   c_{14}\left(u,u_j\right)
   c_{210}\left(u_j,u_k\right)-Y_{112}\left(u,u_j\right) e_{05}\left(u_j,u_k\right)\right],
\end{eqnarray}
\begin{eqnarray}
&&Q_{21}^{\mathcal{W}}(u,u_j,u_k)=k^{+}_{23}(u)\nonumber\\&\times&
\left[a_{12}\left(u_j,u_k\right) \left(a_{25}\left(u,u_j\right)
   c_{211}\left(u,u_j\right)+a_{35}\left(u,u_j\right)
   c_{213}\left(u,u_j\right)\right)
   \right.\nonumber\\&+&\left.
   a_{21}\left(u_j,u_k\right) \left(a_{24}\left(u,u_k\right)
   c_{212}\left(u,u_j\right)+a_{34}\left(u,u_k\right)
   c_{214}\left(u,u_j\right)\right)
   \right.\nonumber\\&+&\left.
   a_{22}\left(u_j,u_k\right) \left(a_{25}\left(u,u_j\right)
   c_{212}\left(u,u_j\right)+a_{35}\left(u,u_j\right)
   c_{214}\left(u,u_j\right)\right)
   \right.\nonumber\\&+&\left.
   a_{24}\left(u_j,u_k\right) \left(c_{212}\left(u,u_j\right)
   x_{27}\left(u,u_j\right)+c_{214}\left(u,u_j\right)
   x_{37}\left(u,u_j\right)\right)
   \right.\nonumber\\&+&\left.
   c_{24}\left(u,u_j\right)
   \left(c_{211}\left(u_j,u_k\right)+c_{27}\left(u_j,u_k\right)\right)-Y_{215}\left(u,u_j\right)
   e_{04}\left(u_j,u_k\right)\right]
   \nonumber\\&+&
   k^{+}_{12}(u) \left[a_{25}\left(u,u_j\right) c_{110}\left(u,u_j\right)
   a_{12}\left(u_j,u_k\right)
   \right.\nonumber\\&+&\left.
   c_{111}\left(u,u_j\right) \left(a_{24}\left(u,u_k\right)
   a_{21}\left(u_j,u_k\right)+a_{25}\left(u,u_j\right)
   a_{22}\left(u_j,u_k\right)+x_{27}\left(u,u_j\right)
   a_{24}\left(u_j,u_k\right)\right)
   \right.\nonumber\\&+&\left.
   c_{14}\left(u,u_j\right)
   \left(c_{211}\left(u_j,u_k\right)+c_{27}\left(u_j,u_k\right)\right)-Y_{113}\left(u,u_j\right)
   e_{04}\left(u_j,u_k\right)\right],
\end{eqnarray}
\begin{eqnarray}
&& Q_{22}^{\mathcal{W}}(u,u_j,u_k)=k^{+}_{23}(u)\nonumber\\&\times&
\left[a_{13}\left(u_j,u_k\right) \left(a_{25}\left(u,u_j\right)
   c_{211}\left(u,u_j\right)+a_{35}\left(u,u_j\right)
   c_{213}\left(u,u_j\right)\right)
   \right.\nonumber\\&+&\left.
   a_{25}\left(u,u_k\right) c_{212}\left(u,u_j\right)
   a_{21}\left(u_j,u_k\right)+a_{35}\left(u,u_k\right) c_{214}\left(u,u_j\right)
   a_{21}\left(u_j,u_k\right)
   \right.\nonumber\\&+&\left.
   a_{23}\left(u_j,u_k\right) \left(a_{25}\left(u,u_j\right)
   c_{212}\left(u,u_j\right)+a_{35}\left(u,u_j\right)
   c_{214}\left(u,u_j\right)\right)
   \right.\nonumber\\&+&\left.
   c_{212}\left(u,u_j\right) x_{27}\left(u,u_j\right)
   a_{25}\left(u_j,u_k\right)+c_{214}\left(u,u_j\right) x_{37}\left(u,u_j\right)
   a_{25}\left(u_j,u_k\right)
   \right.\nonumber\\&+&\left.
   c_{24}\left(u,u_j\right)
   c_{212}\left(u_j,u_k\right)-Y_{215}\left(u,u_j\right) e_{05}\left(u_j,u_k\right)\right]
   \nonumber\\&+&
   k^{+}_{12}(u)
   \left[a_{25}\left(u,u_j\right) c_{110}\left(u,u_j\right)
   a_{13}\left(u_j,u_k\right)
   \right.\nonumber\\&+&\left.
   c_{111}\left(u,u_j\right) \left(a_{25}\left(u,u_k\right)
   a_{21}\left(u_j,u_k\right)+a_{25}\left(u,u_j\right)
   a_{23}\left(u_j,u_k\right)+x_{27}\left(u,u_j\right)
   a_{25}\left(u_j,u_k\right)\right)
   \right.\nonumber\\&+&\left.
   c_{14}\left(u,u_j\right)
   c_{212}\left(u_j,u_k\right)-Y_{113}\left(u,u_j\right) e_{05}\left(u_j,u_k\right)\right].
\end{eqnarray}

We do not present the explicit expressions
for the polynomials $Q_{jk\ell}^{\mathcal{X},\mathcal{Y},\mathcal{Z}}$
since they are very cumbersome and are not necessary in the determination of the $g-$coefficients.
An additional systematic analysis of the commutation relations may allow us to writte these
coefficients in a more manageable form.

\end{document}